\documentclass[aps,pra,twocolumn,showpacs,showkeys,nofootinbib]{revtex4-1}
\usepackage{comment}
\usepackage{enumerate}
\usepackage{tipa}
\usepackage{bbold}
\usepackage{graphicx}
\usepackage{mathrsfs}
\usepackage{amssymb,bm,mathtools}
\usepackage{hyperref}
\usepackage{enumerate}
\usepackage{framed}
\def\I{\mathrm{I}}
\def\II{\mathrm{II}}
\def\Reals{\mathbb{R}}\def\Cmplx{\mathbb{C}}\def\Z{\mathbb{Z}}
\def\>{\rangle}
\def\<{\langle}

\renewcommand{\v}[1]{\ensuremath{\mathbf{#1}}}
 
\newcommand{\ket}[1]{\left| #1 \right>}
\newcommand{\bra}[1]{\left< #1 \right|}
\newcommand{\braket}[2]{\left< #1 \vphantom{#2} \right|
 \left. #2 \vphantom{#1} \right>}

\newtheorem{dfn}{Definition}

\renewcommand{\v}[1]{\ensuremath{\mathbf{#1}}}
\def\I{\mathrm{I}}
\def\II{\mathrm{II}}
\def\Reals{\mathbb{R}}\def\Cmplx{\mathbb{C}}\def\Z{\mathbb{Z}}
\def\>{\rangle}
\def\<{\langle}

\begin{document}
\title{Virtually Abelian Quantum Walks}

\author{Giacomo Mauro D'Ariano}
\author{Marco Erba}
\author{Paolo Perinotti}
\author{Alessandro Tosini}
\affiliation{Universit\`a degli Studi di Pavia, Dipartimento di Fisica, QUIT Group, and INFN Gruppo IV, Sezione di Pavia, via Bassi 6, 27100 Pavia, Italy}

\begin{abstract}
We study discrete-time quantum walks on Cayley graphs of non-Abelian
  groups, focusing on the easiest case of virtually Abelian groups. We
  present a technique to reduce the quantum walk to an
  equivalent one on an Abelian group with coin system having larger
  dimension. This method allows one to extend the notion of wave-vector to the virtually Abelian case and study analytically the walk dynamics. We apply the technique in the case of two quantum walks on
  virtually Abelian groups with planar Cayley graphs, finding the exact solution in terms of dispersion relation.
\end{abstract}
\maketitle

\section{Introduction}

Discrete-time quantum walks (QWs)
\cite{ambainis2001one,aharonov2001quantum,doi:10.1080/09500340408232489}
describe the unitary evolution of a quantum ``particle'' on a
graph. Every node of the graph corresponds to a finite dimensional
Hilbert space---the so-called \emph{coin} system, representing the internal
degree of freedom of the particle.
The graph edges, on the other hand, represent local
interactions among neighbouring sites. 

Under the hypothesis of locality and homogeneity of the evolution one can show \cite{PhysRevA.90.062106} that the graph is the \emph{Cayley graph} of a group.  This allows one to use group representation theory to analyze the dynamics of the particle described by the walk. Classical random walks on Cayley graphs were studied as an integral part of discrete group theory \cite{gromov}, and only later developed as models of computation e.g.~in Ref.~\cite{2012arXiv1212.0027A}. The first paper addressing the issue of constructing QWs on Cayley graphs is Ref.~\cite{acevedo2006quantum}, where the focus is on Cayley graphs of free Abelian groups $\mathbb{Z}^d$. A first analysis investigating QWs on non-Abelian groups can be found in Ref.~\cite{Acevedo:2008:ESQ:2011752.2011757}. Here the authors consider the special class of scalar QWs (the coin system has dimension one) and classify the walks on Cayley graphs of finite
groups presented with two or three generators.

In the present paper we study QWs on Cayley graphs, focusing attention on those groups that can be embedded in a Euclidean manifold quasi-isometrically, i.e.~without distorting the distance defined in terms of minimum arc length on the graph (see Section \ref{s:cayley}). Geometrically, QWs on these groups are the most general QWs with a flat geometry. The class of Cayley graphs that are quasi-isometrically embeddable in a Euclidean manifold coincides with {\em virtually Abelian} groups---including Abelian groups as a special case---that are generally non-Abelian.  A critical issue for QWs on such groups is that, differently from the continuous flat case, they cannot be simply represented in the wave-vector space via the Fourier transform, computing for example the walk \emph{dispersion relation}. Here we bridge this gap, showing how to solve the dynamics for walks on virtually Abelian groups in terms of a wave-vector representation.

The proposed solution consists in a coarse-graining procedure for
coined quantum walks. Tiling procedures for QWs were recently proposed in Refs.~\cite{Portugal2015,PhysRevA.91.052319}, 
where they are exploited for the study of QW search algorithms. The coarse-graining technique that we introduce here 
allows us to reduce the walks on virtually Abelian groups to walks on Abelian groups with a larger coin
system. 

On technical grounds, the above result enables the application of the Fourier analysis technique,
with the definition of plane-waves and the adoption of the wave-vector as
an invariant of the dynamics. As regards the physical motivation for the present analysis, we remind that the theory of quantum walks on Abelian groups has been used in Refs.~\cite{PhysRevA.90.062106,Bisio2015244,Bisio2016177} providing a mechanism for understanding the dynamics of free relativistic quantum fields. In the same perspective, the approach that is proposed in the present paper can provide a similar understanding of the origin of spin and charge degrees of freedom. If the original walk has a trivial coin, indeed, the coarse-grained representation evolves particles of a spinorial or vectorial field, thus carrying non-trivial internal degrees of freedom, as studied in Ref.~\cite{PhysRevA.93.062334}, where the one-dimensional Dirac walk is derived by coarse-graining of a walk with trivial coin on the Cayley graph of the infinite dihedral group.

The technique that we introduce reduces the mathematical
description of the walk to a simpler one, without loss of any
information, and in this respect it cannot be considered a
renormalization technique, in the spirit of Kadanoff's proposal for the
treatment of discrete systems on larger scales \cite{K66}. However,
our approach can also suggest the basic step for implementing the
algebraic procedure of Ref.~\cite{1367-2630-17-8-083005}, where
renormalization is described in terms of CP embedding of C*-algebras
corresponding to the system observables at different scales.

\section{Cayley graphs}\label{s:cayley}
In this section we review the notion of the Cayley graph of a group, and that of quasi-isometries between 
metric spaces, which allows to characterize the class of Cayley graphs quasi-isometrically embeddable in Euclidean spaces.

We can always think of a group $G$ as generated by a set of its elements $S_+$ called generators. We
will denote by $S=S_+\cup S_-$ the union of $S_+$ with the set of its inverses $S_-$. Using $S$ as
an alphabet, we formally build up words $w$ that represent multiplication of generators,
corresponding to the group element $[w]\in G$. Generally there exist different words $w$ such that
$[w]=e\in G$ is the identity. A set of relators $R=\{w_n\}$ is a set of words with $[w_n]=e$ such
that every word $w$ with $[w]=e$ is a juxtaposition of words in $R$. For every group $G$, a choice of
a generating set $S_+$ and a relator set $R$ provides a presentation $G=\<S_+|R\>$ of the group. Any
presentation of $G$ completely specifies $G$, and has the following graphical representation.

\begin{dfn}[Cayley graph]
  Given a group $G$ and a set $S_+$ of generators of the group, the Cayley graph $\Gamma (G, S_+)$
  is defined as the directed edge-colored graph having vertex set $G$, edge set $\{(x, xg); x \in G, g
  \in S_+\}$, and a color assigned to each generator $g \in S_+$.
\end{dfn}

\begin{figure}
\includegraphics[width=.4\linewidth]{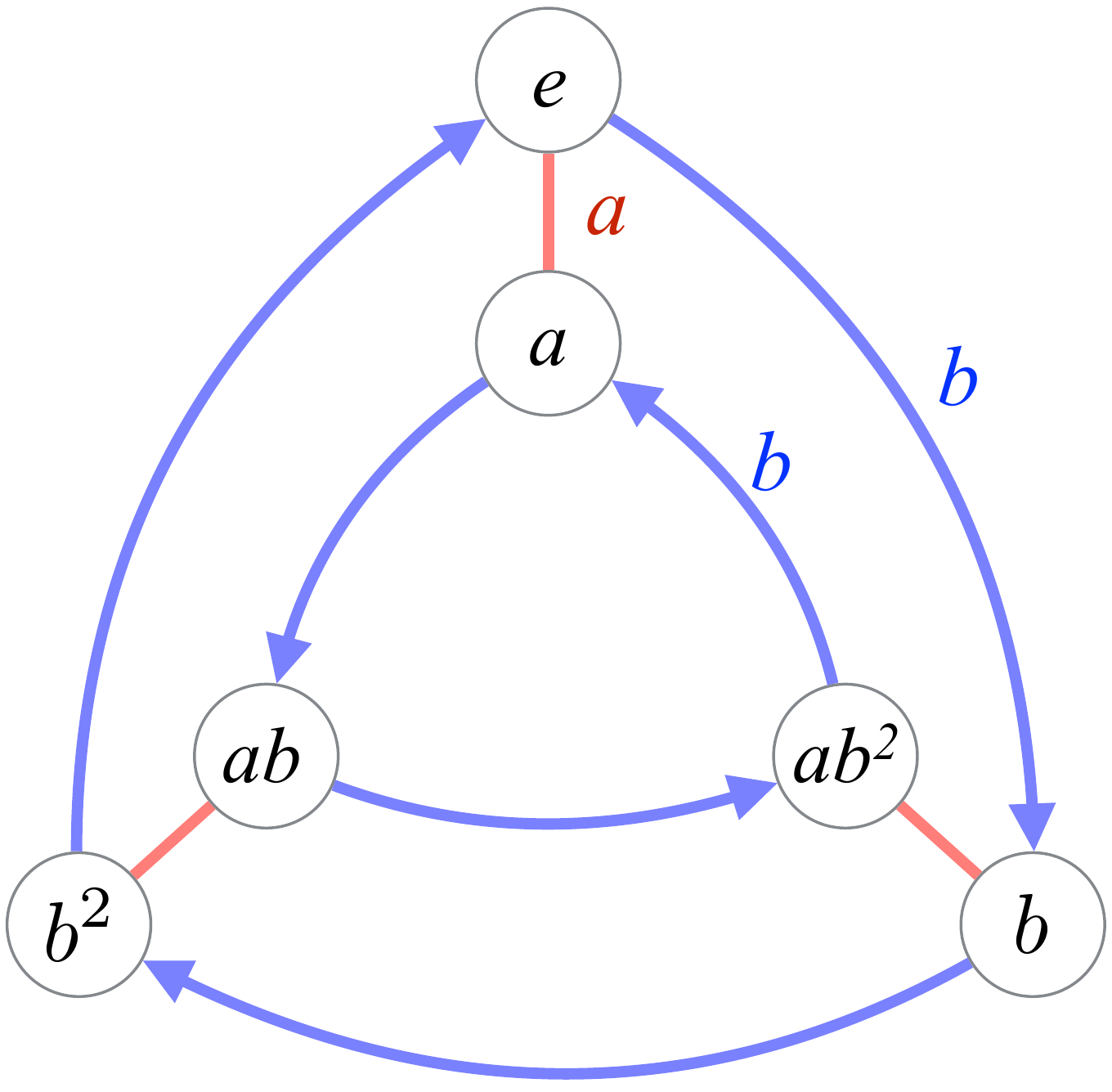}\qquad
\includegraphics[width=.9\linewidth]{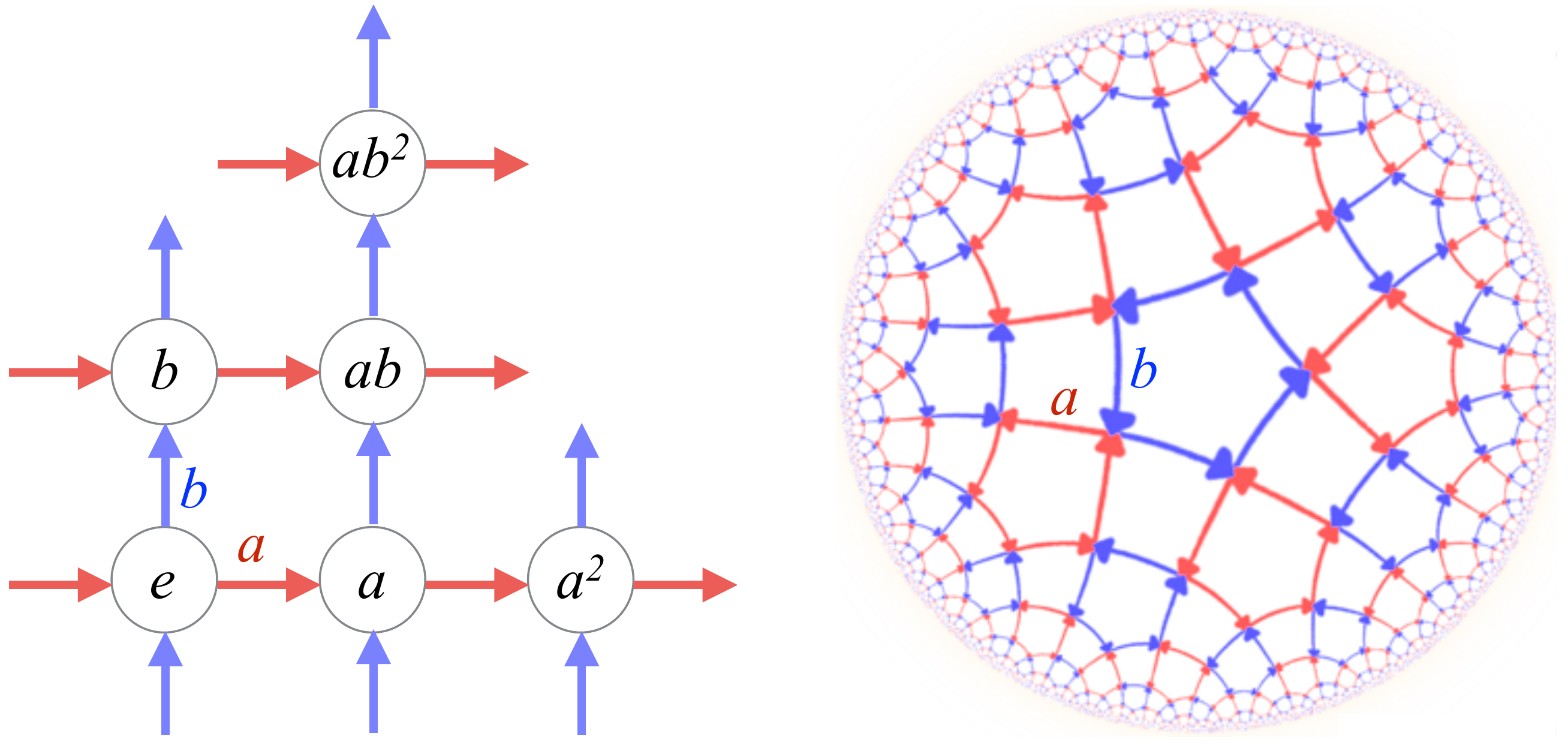}
\caption{(colors online)  Given a group $G$ and a set $S_+$ of
     generators, the Cayley graph $\Gamma(G,S_+)$ is defined as the
    colored directed graph having set of nodes $G$, set of edges
   $\{(x,xh); x \in G, h \in S_+\}$, and a color assigned to each
  generator $h \in S_+$. Top: the Cayley graph of the dihedral group $D_3=\langle a,b | a^2,b^3, (ab)^2 \rangle$.  Bottom left: the Cayley graph of the Abelian
     group $\Z^2$ with presentation $\Z^2=\langle a,b|aba^{-1}b^{-1}\rangle$, where $a$ and $b$   are two commuting generators. Bottom right: the Cayley graph of
      the non-Abelian group $G=\langle a,b|a^5,b^5,(ab)^2\rangle$.
     The Abelian-group graph is embedded into the Euclidean space
    $\mathbb{R}^2$, the non-Abelian $G$ into the Hyperbolic space
        $\mathbb{H}_2$ with (negative) curvature. 
}\label{fig:cayley} 
\end{figure}

In the above representation the relators of the group are closed paths over the graph.\\
In the left of Fig.~\ref{fig:cayley} we see an example of Cayley graph
for the finite dihedral group $G=D_3$ with presentation
$D_3=\langle a,b | a^2,b^3, (ab)^2 \rangle$. In the middle and right
of Fig.~\ref{fig:cayley} we show two examples of Cayley graphs for
infinite groups: the Abelian $\Z^2=\<a,b|aba^{-1}b^{-1}\>$ and
the non-Abelian group $F=\langle a,b|a^5,b^5,(ab)^2\rangle$ (a special
case of Fuchsian group).

Defining the {\em length} $l(w)$ of a word $w$ as the number of
generators that compose it, one introduces the {\em word-distance} of
two points $x_1,x_2\in G$ as the length of the shortest word $w$ such
that $[w]=x_1x_2^{-1}$. The Cayley graph equipped with the word-metric
is a metric space.
In the following section the Cayley graphs will be the lattice of a quantum walk. 

Even for finitely presented groups (namely with $|S_+|$ and $|R|$ both
finite), the algebraic properties of the group can be very hard to
assess, or even provably undecidable (this is the case of triviality
$G=\{e\}$, which is one of the Dehn problems
\cite{dehn1911unendliche}). An effective way of studying the algebraic
properties of groups is that of connecting them with the geometry of
their Cayley graphs: this is the aim of {\em geometric group theory}
\cite{harpe}. The main idea of the approach is the notion of {\em
  quasi-isometry} due to Gromov \cite{gromov1984infinite},
which is defined as follows. Given two metric spaces $(M_1,d_1)$,
$(M_2,d_2)$, we say that $M_1$ is quasi-isometric to $M_2$ if there is
a map $f:M_1\to M_2$ such that there exist three constants $a>1$,
$b>0$, $c>0$, such that $\forall x,y\in M_1$ one has
\begin{align*}
\frac 1a d_2[f(x),f(y)]-b\leq d_1(x,y)\leq a d_2[f(x),f(y)]+b,
\end{align*}
and
$\forall z\in M_2$, there exists $x\in M_1$ such that
\begin{align*}
d_2[z,f(x)]\leq c.
\end{align*}

The quasi-isometry is an equivalence relation between metric
spaces \cite{campbell1999groups}. For example, all Cayley graphs of a group $G$ are
quasi-isometric \cite{campbell1999groups}, and we will transfer the quasi-isometric class to $G$
itself.  Moreover, quasi-isometric groups share relevant algebraic
features. For example a group $G$ is quasi-isometric to a
Euclidean space $\Reals^n$ if and only if it is \emph{virtually Abelian}
\cite{Kapovich:2012te,rigidity}, namely it has an Abelian subgroup $H$ with finite
{\em index} (the number of cosets). In such a case $H\subseteq G$ is
isomorphic to $\Z^n$ with $n\geq 0$. Two examples of virtually Abelian
groups are considered in detail in Sections \ref{s:example1} and
\ref{s:example2} (see also Figs.~\ref{fig:example1} and
\ref{fig:example2}).
Therefore, if $G$ is not virtually Abelian and embeddable in a metric manifold, then the manifold must have a
nonvanishing curvature. An example is provided in the right of Fig.~\ref{fig:cayley}.

\section{Quantum walks on Cayley graphs}\label{s:qws-on-cayley}

A quantum walk with $s$-dimensional coin system ($s\geq 1$) is a unitary evolution of a system with Hilbert space
$\ell^2(V)\otimes\mathbb{C}^s$ such that
\begin{align*}|\psi_{x,t+1}\>=\sum_{y\in V}A_{x,y}|\psi_{y,t}\>,
\end{align*}
where $V\subseteq\Z$ and $A_{x,y}\in M_s(\mathbb{C})$ for $x,y\in V$
are $s\times s$ {\em transition matrices}. The set of nonnull matrices $A_{x,y}$
define a set $E\subseteq V\times V$ which can be regarded as the edge
set of a graph $\Gamma(V,E)$. As we will see in the following, the
conditions for the evolution to be unitary correspond to nontrivial
constraints for the set of transition matrices.  The case of interest
in the present paper is when the graph is the Cayley graph
$\Gamma(G,S_+)$ of a group $G$, where the elements of $G$ are in bijective correspondence with the graph vertices $x\in V$. 
In addition the different colors and orientations of the edges from $x\in V$ correspond to
different transition matrices $\{A_g\}_{g\in S}$ (meaning that
$\forall x\,A_{x, xg}=A_g$). The independence of the matrix set
of $x\in G$ corresponds to the homogeneity of the walk, whereas the
finiteness of the set $S$ is the assumption of locality
\cite{PhysRevA.90.062106}.

If we now consider the right regular representation $T_g$ of the finitely generated group $G$,
acting on $\ell ^2(G)$ as $T_g|x\>:=|xg^{-1}\>$, we can write the quantum walk as
\begin{align}\label{eq:qw}
A:=\sum_{g\in S}T_{g}\otimes A_g.
\end{align}
The unitarity conditions for the walk operator \eqref{eq:qw} are
$AA^{\dagger}=A^{\dagger}A=T_e\otimes I_s$ which imply
\begin{align}\label{eq:unitarity-ort}
 \sum\limits_{gg'^{-1}=f} A_gA_{g'}^{\dagger} &=
  \sum\limits_{g^{-1}g'=f} A_g^{\dagger}A_{g'}=0, \ \ \ \text{(with
    $f\neq e$)},\\ \label{eq:unitarity-norm}
 \sum\limits_{g\in S} A_gA_{g}^{\dagger} &= \sum\limits_{g\in S}
  A_g^{\dagger}A_{g}=I_s,
\end{align}
where $I_s$ is the identity operator over $\Cmplx^s$.
Another relevant symmetry of a QW on a Cayley graph is the \emph{isotropy}, saying that ``any
direction on $\Gamma(G,S_+)$ is equivalent''. This condition is translated in mathematical terms
requiring that there exists a subset $S'$ of $S$, with $S_+\subseteq S'$, and a unitary representation $U$ over $\mathbb{C}^s$ of
a group $Q$ of graph automorphisms, transitive over $S'$, such that one has the covariance
condition
\begin{equation}\label{eq:isotropy}
A=\sum_{g\in S} T_g\otimes A_g=\sum_{g\in S} T_{f(g)}\otimes U_fA_gU^\dag_f,\quad \forall f\in Q.
\end{equation}
As shown in \cite{PhysRevA.90.062106}, QWs satisfying isotropy can be classified imposing the condition
\begin{equation}\label{eq:isotropy2}
\sum\limits_{g\in S}A_g =I_s
\end{equation}
and then multiplying the resulting transition matrices by an arbitrary unitary matrix commuting with the representation $U$ of the automorphisms group.

\subsection{Quantum walks on Cayley graphs of Abelian groups}

The simplest case of quantum walks on Cayley graphs is when the group is free Abelian, i.e. $\Z^d$,
since the walk can be easily diagonalized by a Fourier transform. We will label the elements
$\v{x}\in\Z^d$, using the usual additive notation for the group composition. The right regular
representation of $\Z^d$ is expressed as follows
\begin{equation*}
T_{\v{y}}\ket{\v{x}}=\ket{\v{x}-\v{y}}.
\end{equation*}
The representation is decomposed into irreducible representations, which are all one-dimensional, over which one
diagonalizes $T_{\v{y}}$ via Fourier in the wave-vector space as follows
\begin{equation*}
\ket{\v{k}}\mathrel{\mathop:}=\frac{1}{\left(2\pi\right)^{\frac{d}{2}}}\sum\limits_{\v{x}\in \Z^d} e^{-i\v{k}\cdot \v{x}}\ket{\v{x}},
\end{equation*}
where $\v k$ belongs to the {\em first Brillouin zone} $\mathcal{B}\subseteq \mathbb R^d$, which is the largest set 
that contains vectors $\v k$ corresponding to inequivalent elements $\ket{\v k}$.
The walk is written in the direct integral decomposition
\begin{align*}
A= \int_{\mathcal{B}} d\v{k} \ket{\v{k}}\bra{\v{k}}\otimes A_{\v{k}},\quad A_{\v{k}}:=\sum_{\v{h}\in S} e^{-i\v{k}\cdot \v{h}}A_{\v{h}},
\end{align*}
where $A_{\v{k}}$ is unitary for every $\v{k}\in \mathcal{B}$.
Notice that the Brillouin zone depends on $S$, which in turn corresponds to a specific Cayley graph of $\Z^d$; e.g.
for $\mathbb Z^3$ one has different Brillouin zones depending on the choice of presentation, which can correspond 
to the simple cubic lattice, or the body-centered cubic one, etc. (for details, see Ref. \cite{PhysRevA.90.062106}).\\
Diagonalizing $A_{\v{k}}$, one
obtains
\begin{align*}
A_{\v{k}}\ket{u_r(\v k)}=e^{i \omega_r (\v{k})}\ket{u_r(\v k)},
\end{align*}
where the $\omega_r (\v{k})$ are the \emph{dispersion relation} of the walk, and
$\ket{u_r(\v k)}$ the corresponding eigenstates. The dispersion relation gives the kinematics of the
walk, with its first and second derivatives providing respectively the \emph{group velocity} and the
\emph{diffusion coefficient} of particle states.

While for free Abelian groups the Fourier transform approach is straightforward, this is no longer
true for generally non-Abelian groups. However, for virtually Abelian groups the procedure is still
possible as we will see in the next section. 

\section{Virtually Abelian quantum walks}\label{s:reg}

In this section we study quantum walks on Cayley graphs of a virtually
Abelian group $G$ with finite-index Abelian subgroup $H\cong \Z^d$ (this
choice is not restrictive by the classification of Abelian groups). We
will see how the virtual Abelianity of the group allows one to define
a coarse-graining of the Cayley graph that leads to an equivalent
quantum walk on a Cayley graph of $H$. The basic idea is to partition
$G$ into cosets of $H$, denoting them by a finite set of labels. 
In the following we will consider right cosets
without loss of generality. The vertices of the Cayley graph of $G$
are thus grouped into clusters---containing one vertex from each coset---that become the nodes of a new
coarse-grained graph and are in correspondence with the elements of $H$.  
Correspondingly, one can find an equivalent walk in terms of the generators of $H$, designating the coset labels as additional internal degrees of freedom.

Firstly we will provide a formal definition of the intuitive notion of ``regular tiling of a Cayley graph''.

\begin{dfn}[Regular tiling]\label{d:regular-tiling}
  Let $G$ be a virtually Abelian group and let $H$ be an Abelian subgroup of $G$ of finite index. If
  $G$ is finitely generated and $H\cong \Z^d$, we call {\em regular tiling of order $l$} the
  following right cosets partition for the Cayley graph of $G$
\begin{equation*}
\bigcup\limits_{j=1}^{l}Hc_j=G \qquad \text{(with $c_1=e$)}.
\end{equation*}
\end{dfn}
Notice that the regular tiling is not unique and depends on the choice of $H$ and that of the
representatives $c_j$. For any choice we can now provide a wave-vector representation of the quantum
walk.

Consider the quantum walk $A$ on the Cayley graph given by the following presentation
\begin{equation}\label{eq:walk1}
\begin{aligned}
&G:=\langle S_+|R\rangle=\langle g_1,\ldots g_n|r_1,\ldots, r_m\rangle,\\
&A:=\sum_{g\in S}T_g\otimes A_g,\quad \left(A:\ell ^2(G)\otimes\mathbb
  C^s\rightarrow\ell^2(G)\otimes\mathbb C^s\right). 
\end{aligned}
\end{equation}
Consider a regular tiling for $G$ corresponding to an Abelian subgroup $H$ of index $l$ and right
cosets representatives $c_1,\ldots,c_l$. We show how the virtually Abelian quantum walk
\eqref{eq:walk1} on $\ell ^2(G)\otimes\mathbb C^s $ can be regarded as an Abelian QW on $\ell
^2(H)\otimes\mathbb C^{sl}$.

In the following we will adopt the vector notation for the elements of $H$, keeping the multiplicative 
notation for the general group composition in $G$. While this choice may be slightly confusing, it is 
unavoidable to define the Fourier transform. The notation $\v{x}c_j$ should thus not be interpreted 
as the rescaling of the vector $\v x$ by a scalar, but just as group composition in $G$.

From the disjointness of the cosets it follows that every $x\in G$ admits the unique decomposition
$x=\v{x}c_j$ with $\v{x}\in H$. One can then define a unitary mapping
between $\ell^2\left(G\right)$ and $\ell^2\left(H\right)\otimes\mathbb{C}^l$ as follows
\begin{align*}
U_H:\ell^2\left(G\right) &\longrightarrow\ell^2\left(H\right)\otimes\mathbb{C}^l,\qquad
     U_H\ket{\v{x}c_j}=\ket{\v{x}}\ket{j}.
\end{align*}
Now we can define the plane waves on cosets
\begin{equation*}
\ket{\v{k}}_j\mathrel{\mathop:}=\frac{1}{\left(2\pi\right)^{\frac{d}{2}}}\sum\limits_{\v{x}\in H} e^{-i\v{k}\cdot \v{x}}\ket{\v{x}c_j},
\end{equation*}
which via the map $U_H$ define the vector $\ket{\v{k}}_{H}$ as follows
\begin{align*}
\begin{split}
U_H\ket{\v{k}}_j & = \ket{\v{k}}_{H}\ket{j}.
\end{split}
\end{align*}
One has that $\forall g\in S$, $\forall \v{x} \in H$ and $\forall c_j$ there exist an element $\v{x'}\in H$ and a coset representative $c_{j'}$ (with $j'$ function of $\left(g,j\right)$) such that
\begin{align*} 
\v{x}c_jg^{-1}=:\v{x}'c_{j'\left(g,j\right)},\qquad
\v{h}_{j,g}\mathrel{\mathop:}=\v{x}-{\v{x}'}=c_{j'\left(g,j\right)}gc_{j}^{-1}\in H.
\end{align*}
Thus one has that the translation operator $T_g$ for an arbitrary generator $g\in S$ of $G$ acts on the coset plane waves as
\begin{align}\label{e:old-action}
\begin{split}
T_g\ket{\v{k}}_j  &=\frac{1}{\left(2\pi\right)^{\frac{d}{2}}}\sum\limits_{\v{x}\in H} e^{-i\v{k}\cdot \v{x}}\ket{\v{x}c_jg^{-1}}  = \\
&= \frac{1}{\left(2\pi\right)^{\frac{d}{2}}}\sum\limits_{\v{x}'\in H} e^{-i\v{k}\cdot \left(\v{h}_{j,g}+\v{x}'\right)}\ket{\v{x}'c_{j'\left(g,j\right)}} = \\
&= e^{-i\v{k}\cdot \v{h}_{j,g}}\ket{\v{k}}_{j'\left(g,j\right)},
\end{split}
\end{align}
and on the wave-vector $\ket{\v{k}}_{H}\ket{j}$ as
\begin{align}\label{e:new-action}
\begin{split}
\left(U_HT_gU_H^{\dagger}\right) \ket{\v{k}}_{H}\ket{j} & =
e^{-i\v{k}\cdot\v{h}_{j,g}}\ket{\v{k}}_H\ket{j'\left(g,j\right)}. 
\end{split}
\end{align}
The last equation shows that $\ket{\v{k}}_{H}$ is an invariant space of the {\em coarse-grained generators}
\begin{align}
\tilde{T}_g\mathrel{\mathop:}=U_HT_gU_H^{\dagger} \quad(\tilde{T}_g: \ell^2(H)\otimes\mathbb{C}^l\rightarrow \ell^2(H)\otimes\mathbb{C}^l).
\end{align}
While $T_g$ in the original walk description was changing the coset wave-vector (see
Eq.~\eqref{e:old-action}), its coarse-grained version $\tilde{T}_g$ keeps the wave-vector
$\ket{\v{k}}_{H}$ invariant (up to a phase factor dependent on the generator $g$) with a non
trivial action only on the additional coin system $\mathbb{C}^l$ (see Eq.~\eqref{e:new-action}).

Accordingly, this mapping allows to diagonalize the generators of $G$ over the wave-vector space of
$\ell^2(H)\otimes\mathbb{C}^l$ exploiting their action on the additional degrees of freedom. The
coarse-grained quantum walk is thus given by
\begin{align*}
\mathcal{R}\left[ A\right]\mathrel{\mathop:}=\left( U_H\otimes I\right) A\left( U_H\otimes I\right)^{\dagger}=\sum\limits_g \tilde{T}_g\otimes A_g.
\end{align*}
Now, evaluating the matrix elements for the translations $\tilde T_g$, with $g \in S$
\begin{align*}
\bra{\v{k}}_H\bra{i}\tilde{T}_g \ket{\v{k}}_H\ket{j}=e^{-i\v{k}\cdot\v{h}_{j,g}}\delta_{i,j'\left(g,j\right)},\quad i,j=1,\dots, l,
\end{align*}
we can finally obtain the wave-vector representation of the walk $A_{\v k}$. The last one is
block-partitioned with block $ij$ given by
\begin{equation*}
(A_\v k)_{ij}= \sum_{g:j'(g,j)=i}A_ge^{-i\v k\cdot\v h_{j,g}}.
\end{equation*}

Notice that coarse-grained QWs corresponding to different choices of the regular tiling are unitarily equivalent. Let $U_H,V_{H'}$ be the coarse-graining unitary operators corresponding to two different choices of the subgroup of finite index (or even just to two different choices of the coset representatives for the same subgroup $H=H'$): then, defining $\mathcal{R}\left[ A\right]\mathrel{\mathop:}=\left( U_H\otimes I\right) A\left( U_H\otimes I\right)^{\dagger}$ and $\mathcal{R}'\left[ A\right]\mathrel{\mathop:}=\left( V_{H'}\otimes I\right) A\left( V_{H'}\otimes I\right)^{\dagger}$, one has that the two coarse-grained QWs are connected through the unitary mapping
\[
\mathcal{R}'\left[ A\right]=\left( V_{H'}U_H^{\dagger}\otimes I\right) \mathcal{R}\left[ A\right]\left( V_{H'}U_H^{\dagger}\otimes I\right)^{\dagger}.
\]
In particular, two different choices do not change the dispersion relation.

While we defined the coarse-graining of a QW restricting to the case of interest of virtually Abelian groups, the procedure is general and can be easily defined for QWs on any arbitrary group $G$ which is virtually $G'$ (namely any $G$ having a non-Abelian subgroup $G'$ of finite index $l'$). Indeed, also in this case the procedure can be performed as described above through a unitary mapping $U_{G'}$
between $\ell^2\left(G\right)$ and $\ell^2\left(G'\right)\otimes\mathbb{C}^{l'}$, leading to a QW on a Cayley graph of $G'$ with a larger coin system. However, one is not able to define plane waves on the cosets of $G'$ if this is not Abelian, and this does not allow one to represent the walk in the k-space.

In the following we present the first two examples of QWs on non-Abelian groups whose evolution is analytically solved via the  procedure described in this section.

\subsection{Example of a massive virtually Abelian QW}\label{s:example1}

\begin{figure}
  \includegraphics[width=.95\linewidth]{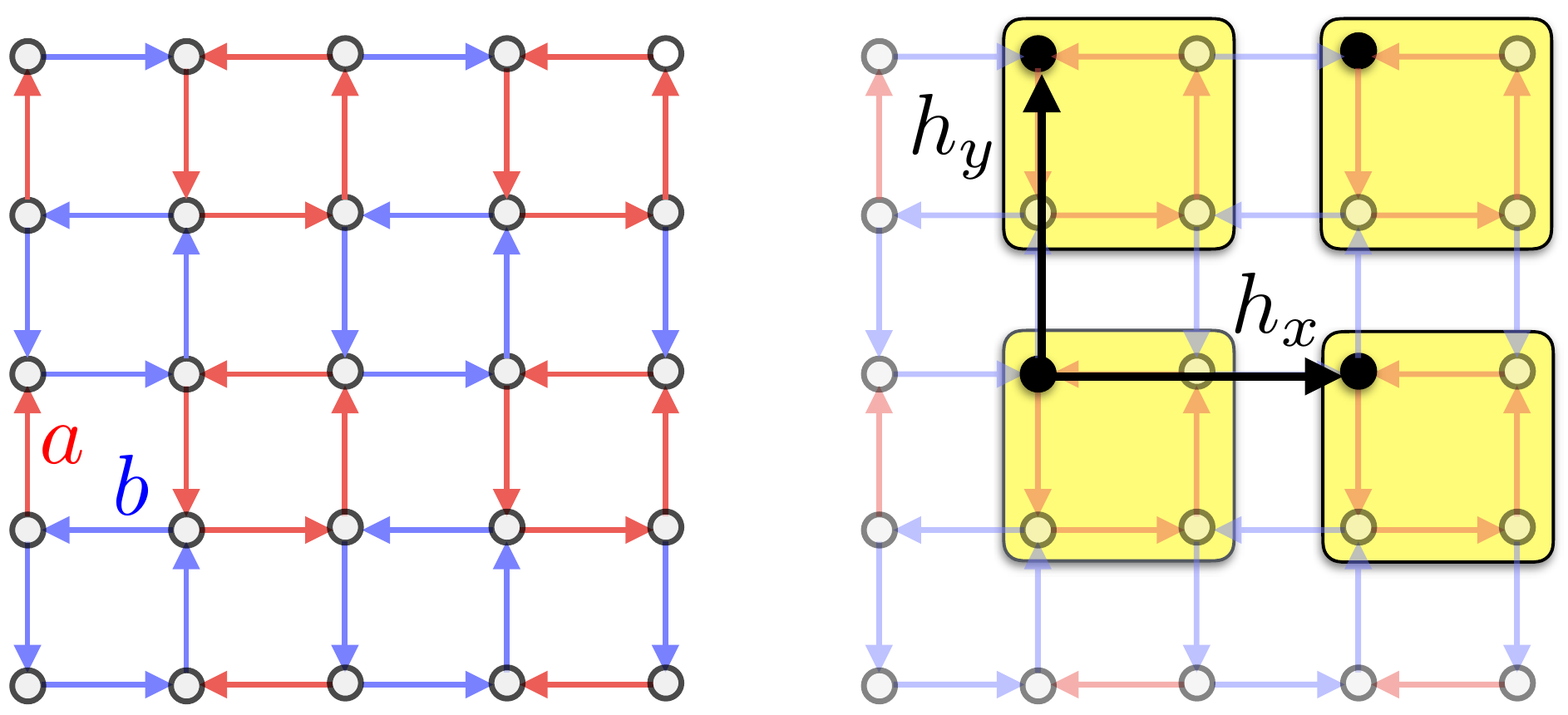}
\caption{(colors online) Left: the Cayley graph of $G_1$
  corresponding to the presentation
  $G_1 = \langle a,b\ |\ a^4,b^4,{\left(ab\right)}^{2}\rangle $. The
  group is clearly not Abelian ($a$ and $b$ do not commute) but it is
  virtually Abelian. Right: a possible regular tiling of the Cayley
  graph of $G_1$. First we choose a subgroup (subgraph) isomorphic to
  $\Z^2$, in the figure the black sites given by the subgroup
  generators $h_x=a^{-1}b$ and $h_y=ba^{-1}$, which commute. The
  subgroup has index four and we choose a set of cosets
  representatives, in the figure $e,a,a^2,a^3$, which define the
  tiling, each tile containing four elements of
  $G_1$.}\label{fig:example1}
\end{figure}

Let us consider the virtually Abelian group $G_1$ with the presentation
\begin{equation*}
  G_1 = \langle a,b\ |\ a^4,b^4,(ab)^2\rangle,
\end{equation*}
and with Cayley graph corresponding to the simple square lattice,
modulo a suitable re-coloring and re-orienting of the edges (see
Fig.~\ref{fig:example1}).

Solving the unitarity constraints for
\begin{equation*}
A=\sum_{g\in\{a,b,a^{-1},b^{-1}\}}T_g\otimes A_g
\end{equation*}
on the Hilbert space $\ell^2(G_1)\otimes \mathbb{C}^s$ we obtain the
admissible QWs on the Cayley graph of $G_1$ and with two-dimensional
coin system. Due to the relators of the $G_1$ the first unitary
condition \eqref{eq:unitarity-ort} leads to the following relations
\begin{align}\begin{split}\label{eq:ort1}
 & A_{i^{-1}}A_{{j}^{-1}}^{\dagger} =
    A_{i^{-1}}^{\dagger}A_{{j}^{-1}} =\\
&= A_iA_{{j}}^{\dagger}  =
    A_i^{\dagger}{A_{{j}}} = 0, \qquad   i\neq j\in\{a,b\}\end{split} \\ 
    \begin{split}\label{eq:sum1}
  &  A_iA_{j^{-1}}^{\dagger}+A_{{j}^{-1}}A_i^{\dagger}  = \\
  &= A_{i^{-1}}^{\dagger}A_j+A_j^{\dagger}A_{{i}^{-1}}
    = 0,\qquad i,j\in\{a,b\},
    \end{split}
\end{align}    
while the constraint
\eqref{eq:unitarity-norm} gives
\begin{align}\begin{split}\label{eq:norm1}
  & A_aA^\dag_a+A_bA^\dag_b+A_{a^{-1}}A^\dag_{a^{-1}}+A_{b^{-1}}A^\dag_{b^{-1}}= \\
  &=A^\dag_aA_a+A^\dag_bA_b+A^\dag_{a^{-1}}A_{a^{-1}}+A^\dag_{b^{-1}}A_{b^{-1}}=I_s.
\end{split}\end{align}
From Eqs. (\ref{eq:ort1}) one can see that for $s=1$ at least one of
the transition matrices must be null, contradicting the hypothesis
(the edges of the graph correspond to non null transition matrices)
hence there is no quantum walk on the considered Cayley graph.  The
simplest case is thus $s=2$, and in Appendix \ref{app:a} we show that assuming
the isotropy of the QW (see Section \ref{s:qws-on-cayley}) the
solutions are divided into two non-unitarily equivalent classes
\begin{align*}
  A^{\I}_a&= \zeta^{\pm} Z_\pm\begin{pmatrix}
    1       &   0  \\
    0    &   0\end{pmatrix},  &      
 A^{\I}_b &= \zeta^{\pm} Z_\pm\begin{pmatrix}
    0       &   0  \\
    0    &   1\end{pmatrix}, \\
  A^{\I}_{a^{-1}} &= (A^{\I}_a)^{\dagger}, &
A^{\I}_{b^{-1}} &=(A^{\I}_b)^{\dagger},
\end{align*}
\begin{align*}
  A^{\II}_a&= A^{\I}_a,&         
 A^{\II}_b &= A^{\I}_b,&
  A^{\II}_{a^{-1}} &= (A^{I}_b)^{\dagger},& 
A^{\II}_{b^{-1}} &=(A^{\I}_a)^{\dagger},
\end{align*}
where $\zeta^{\pm}:=\frac{1 \pm i}{2}$ and $Z_\pm:=nI_2\pm
im\sigma_x$ with $n^2+m^2=1$, $n,m\geq 0$.

In order to solve this class of walks analytically we apply the algorithm of Section \ref{s:reg}.
First we choose an Abelian subgroup of finite index $H\cong \Z^2$, and this is the group generated
by $\v{h}_x=a^{-1}b$, $\v{h}_y=ba^{-1}$ of index four. Then we chose the regular tiling of the
Cayley graph of $G_1$ achieved by the coset partition
\begin{align*}
G_1= \bigcup_{j=0}^3 Hc_j,\quad c_j=a^j.
\end{align*}
Now we can define the plane waves on the cosets
\begin{equation*}
\ket{\v{k}}_j\mathrel{\mathop:}=\frac{1}{2\pi}\sum\limits_{\v{x}\in H}e^{-i\v{k}\cdot \v{x}}\ket{\v{x}a^{j}},\ \ j=0,1,2,3,
\end{equation*}
and compute the action of the original walk on the cosets wave-vectors
\begin{align*}
T_a\ket{\v{k}}_0 &= \ket{\v{k}}_3, &
 T_a\ket{\v{k}}_1 &= \ket{\v{k}}_0, \\
 T_a\ket{\v{k}}_2 &= \ket{\v{k}}_1, &
T_a\ket{\v{k}}_3 &= \ket{\v{k}}_2,\\
T_b\ket{\v{k}}_0 &= e^{-ik_x}\ket{\v{k}}_3, &
T_b\ket{\v{k}}_1 &= e^{-ik_y}\ket{\v{k}}_0, \\
T_b\ket{\v{k}}_2 &= e^{ik_x}\ket{\v{k}}_1, &
T_b\ket{\v{k}}_3 &= e^{ik_y}\ket{\v{k}}_2,\\
T_{a^{-1}}\ket{\v{k}}_0 &= \ket{\v{k}}_1, & 
T_{a^{-1}}\ket{\v{k}}_1 &= \ket{\v{k}}_2, \\
T_{a^{-1}}\ket{\v{k}}_2 &= \ket{\v{k}}_3, &
T_{a^{-1}}\ket{\v{k}}_3  &= \ket{\v{k}}_0, \\
T_{b^{-1}}\ket{\v{k}}_0  &= e^{ik_y}\ket{\v{k}}_1, &
T_{b^{-1}}\ket{\v{k}}_1  &= e^{-ik_x}\ket{\v{k}}_2, \\
T_{b^{-1}}\ket{\v{k}}_2  &= e^{-ik_y}\ket{\v{k}}_3, &
T_{b^{-1}}\ket{\v{k}}_3  &= e^{ik_x}\ket{\v{k}}_0,
\end{align*}
where $k_x=\v{k}\cdot \v{h}_x,\ k_y=\v{k}\cdot \v{h}_y$.
It follows the expression for the coarse-grained QW
\begin{align*}
\mathcal{R}\left[A\right]&=\int_{\mathcal{A}} d\v{k}
  \ket{\v{k}}\bra{\v{k}}_H\otimes
  \mathcal{R}\left[A\right]_{\v{k}},
\end{align*}
where
\begin{align*}
\begin{aligned}
 \mathcal{R}\left[A\right]_{\v{k}} &= \begin{pmatrix}
  A_{k_y}       &    A_{k_x}^{'} \\
  A_{-k_x}^{'}    & A_{-k_y}
\end{pmatrix}\\
A_{k} & = \begin{pmatrix}
  0      &  A_{a}+e^{-ik}A_{b}   \\
  A_{a^{-1}}+e^{ik}A_{b^{-1}}  &   0
\end{pmatrix}, \\
A_k^{'} & = \begin{pmatrix}
  0      &  A_{a^{-1}}+e^{ik}A_{b^{-1}}   \\
  A_{a}+e^{-ik}A_{b}  &   0
\end{pmatrix}.
\end{aligned}
\end{align*}
By diagonalizing the $8\times 8$ matrix
$\mathcal{R}\left[A\right]_{\v{k}}$ we can finally compute the four
eigenvalues $e^{- i\omega^{\pm}_r(k_x,k_y)}$, $r=1,2$ (each with
multiplicity 2) of the coarse-grained walk with
\begin{equation*}
\begin{aligned}
 &\omega^{\pm}_1=\pm \arccos\alpha(\nu,k_1,k_2) - \pi/4,\quad\omega_2=\omega_1^{\pm} - \pi,\\
&\alpha(\nu,k_1,k_2):=\nu\sqrt{\frac{1}{2}\left( \cos^2\frac{k_x}{2}+\cos^2\frac{k_y}{2} \right)},
\end{aligned}
\end{equation*}
and $\nu=n$ ( respectively $\nu=m$) for the first (respectively second) class of solutions. The $\omega_r$
provides the QW dispersion relations providing the information on the
kinematics of particle states. Close to the minimum of the dispersion
relation, namely for small wave-vectors, one recovers a massive
relativistic dispersion relation, with mass given by $\sqrt{1-\nu^2}$.
The mass is upper bounded by one and at the bound the dispersion
relation becomes flat (this feature is due to unitarity and has
already been pointed out in
Refs.~\cite{PhysRevA.90.062106,mauro2012quantum,Bisio2015244}).  In
this example we observe how the non-Abelianity of the group $G_1$
induces a ``massive'' non trivial dynamics for the coarse-grained QW
on the Abelian group $H$.

\subsection{Example of a massless virtually Abelian QW}\label{s:example2}

\begin{figure}
\includegraphics[width=.95\linewidth]{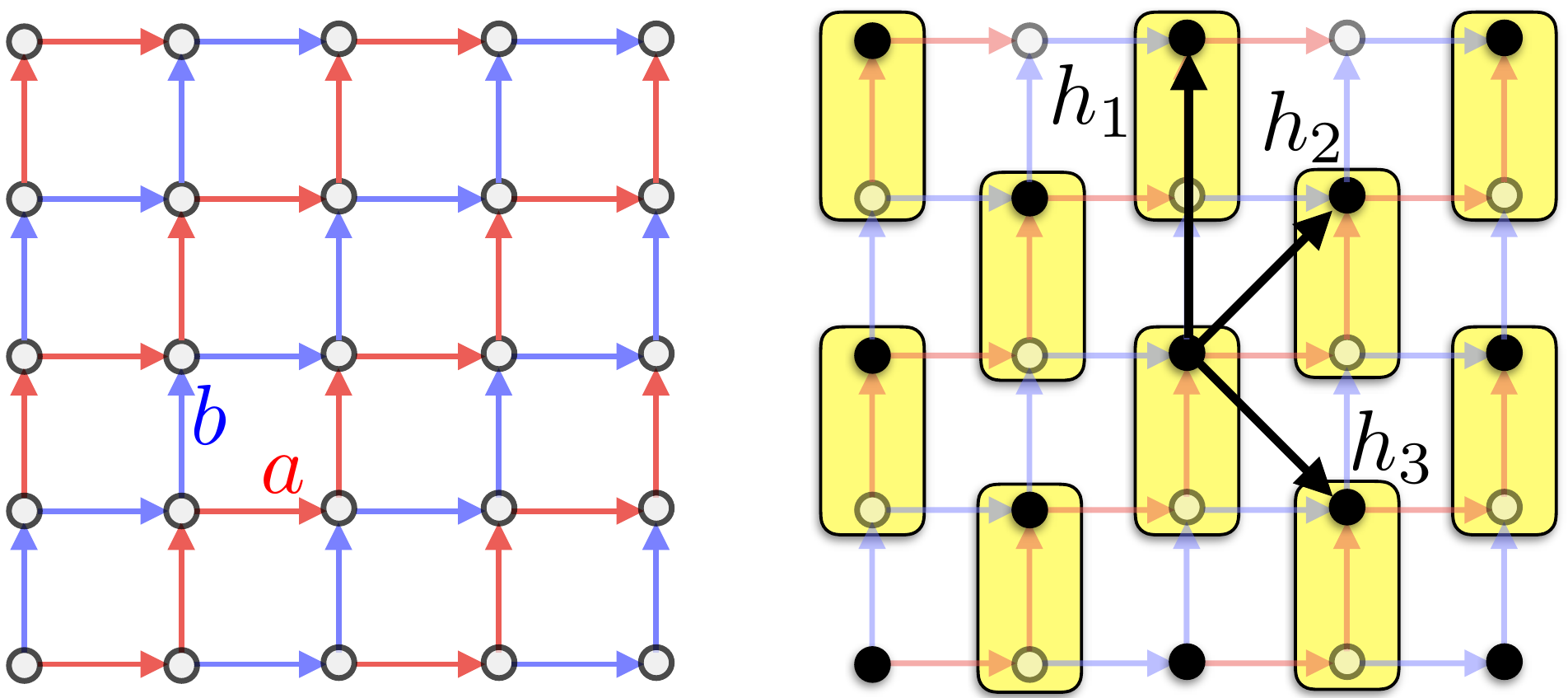}
\caption{(colors online) Left: the Cayley graph of $G_2$
  corresponding to the presentation $G_2 = \langle a,b\ |\
  a^2b^{-2}\rangle  $. The group is not Abelian ($a$ and $b$ do not
  commute) but it is virtually Abelian. Right: a regular tiling of the Cayley graph of $G_2$. First we choose a subgroup (subgraph) isomorphic to $\Z^2$, in the figure the black sites gien by the subgroup commuting generators $h_1=ba$ and $h_2=a^2$. The subgroup has index two and we choose a set of cosets representatives, in the figure $e,a$, which define the tiling, each tile containing two elements of $G_2$.}
\label{fig:example2}
\end{figure}

In this section we consider another virtually Abelian group quasi-isometric
to $\mathbb{R}^2$
\begin{equation*}
G_2 = \langle a,b\ |\ a^2b^{-2}\rangle,
\end{equation*}
whose Cayley graph is depicted in Fig.~\ref{fig:example2} and has vertices
on the simple square lattice.

As in the previous section we start deriving the most general isotropic QW 
\begin{equation*}
A=\sum_{g\in\{a,b,a^{-1},b^{-1}\}}T_g\otimes A_g
\end{equation*}
on the Cayley graph of $G_2$ solving the unitarity constraints on the
transition matrices. Using the relators of the group it is easy
to see that the first unitary condition \eqref{eq:unitarity-ort} leads
to the following relations
\begin{align}\label{eq:ort2}
    & A_{i}A_{{j}^{-1}}^{\dagger} = A_i^{\dagger}A_{{j}^{-1}}= 0,
\\ \begin{split}\label{eq:sum2}
  &  A_iA_j^{\dagger}+A_{{i}^{-1}}A_{{j}^{-1}}^{\dagger}  =  A_i^{\dagger}{A_j}+A_{{i}^{-1}}^{\dagger}{A_{{j}^{-1}}} = \\
 &= A_iA_{{i}^{-1}}^{\dagger}+A_jA_{{j}^{-1}}^{\dagger}  =
    A_i^{\dagger}{A_{{i}^{-1}}}+A_j^{\dagger}{A_{{j}^{-1}}}= 0,\end{split}
\end{align}
for $i\neq j\in\{a,b\}$, while the constraint
\eqref{eq:unitarity-norm} gives
\begin{align}\begin{split}\label{eq:norm2}
  & A_aA^\dag_a+A_bA^\dag_b+A_{a^{-1}}A^\dag_{a^{-1}}+A_{b^{-1}}A^\dag_{a^{-1}}= \\
  &=A^\dag_aA_a+A^\dag_bA_b+A^\dag_{a^{-1}}A_{a^{-1}}+A^\dag_{b^{-1}}A_{b^{-1}}=I_s.
\end{split}\end{align}
As in the previous example there is no QW on the considered Cayley graph
with $s=1$. In Appendix \ref{app:b} we prove that for $s=2$ the only two
solutions are
\begin{align*}
\begin{split}
A^{\I}_a   & = \frac{1}{2} 
\begin{pmatrix}
  1       &   0  \\
  1    &   0
\end{pmatrix},\quad
 A^{\I}_b = \frac{1}{2} 
\begin{pmatrix}
  1       &   0  \\
  -1    &   0
\end{pmatrix}, \\
A^{\I}_{a^{-1}} & = \frac{1}{2}  
\begin{pmatrix}
  0       &   1  \\
  0    &   1  
\end{pmatrix},\quad
A^{\I}_{b^{-1}}= \frac{1}{2}  
\begin{pmatrix}
  0       &  - 1 \\
  0    &   1  
\end{pmatrix}, \end{split}
\\
A_g^{\II} &= Y{\left( A_g^{\I}\right)}^tY^{\dagger},\ \ \
  Y=\frac{1}{\sqrt{2}}\left( I_2 + i\sigma_y \right),
\end{align*}
which are anti-unitarily equivalent.

We choose now the Abelian subgroup $H\cong \Z^2$ of index $2$
generated by $\v{h}_2=a^2$, $\v{h}_3=a^{-1}b$ and the $G_2$ regular tiling partition
\begin{equation*}
G_2= H\cup Hc,\ \ c=a^{-1}.
\end{equation*}
It is then useful to introduce the graph corresponding to the presentation
\begin{equation}\label{H}
H= \langle \v{h}_1,\v{h}_2,\v{h}_3\ |\ \v{h}_1-\left(\v{h}_2-\v{h}_3\right)\rangle ,
\end{equation}
whose vertices are a subset of the vertices of the original Cayley graph of $G$
(the Abelian notation is used since $H\cong \Z^2$) where we added the
redundant generator $\v{h}_1=ba$.

Due to the normality (since it is of index 2) of $H$ in $G_2$ we can define the wave-vector on
cosets, namely the invariant spaces under the $T_{\v{h}_i}$, $i=1,2,3$,
\begin{equation*}
\ket{\v{k}}_0\mathrel{\mathop:}=\frac{1}{2\pi}\sum\limits_{\v{x}\in H}e^{-i\v{k}\cdot \v{x}}\ket{\v{x}},\ \ket{\v{k}}_1\mathrel{\mathop:}=\frac{1}{2\pi}\sum\limits_{\v{x}\in H}e^{-i\v{k}\cdot \v{x}}\ket{\v{x}a^{-1}}.
\end{equation*}
Evaluating the action of the generator of $G_2$ on the
$\ket{\v{k}}_j$, one can reconstruct the coarse-grained walk
$\mathcal{R}[A]$, which will be written in terms of the generators
of $H$ and their inverses. First we evaluate the action of the $G_2$
generators on the cosets wave-vectors
\begin{align*}
T_a\ket{\v{k}}_0  &= \ket{\v{k}}_1,         &  T_a\ket{\v{k}}_1 &= e^{-ik_2}\ket{\v{k}}_0, \\
T_b\ket{\v{k}}_0  &= e^{-ik_3}\ket{\v{k}}_1, &  T_b\ket{\v{k}}_1 &= e^{-ik_1}\ket{\v{k}}_0, \\
T_{a^{-1}}\ket{\v{k}}_0 &= e^{ik_2}\ket{\v{k}}_1, &  T_{a^{-1}}\ket{\v{k}}_1 &= \ket{\v{k}}_0, \\
T_{b^{-1}}\ket{\v{k}}_0  &= e^{ik_1}\ket{\v{k}}_1,  &  T_{b^{-1}}\ket{\v{k}}_1 &= e^{ik_3}\ket{\v{k}}_0.
\end{align*}
It follows the off-diagonal expression for the renormalized walk
\begin{equation}\label{Wna}
\begin{aligned}
\mathcal{R}[A]_{\v{k}} &= \begin{pmatrix}
  0       &  A_{\v{k}} \\
  A_{\v{k}}'      &   0
\end{pmatrix},\\
A_{\v{k}} & = e^{-ik_2}A_a+e^{-ik_1}A_b+A_{a^{-1}}+e^{ik_3}A_{b^{-1}}, \\
A_{\v{k}}' & = A_a+e^{-ik_3}A_b+e^{ik_2}A_{a^{-1}}+e^{ik_2}A_{b^{-1}}.
\end{aligned}
\end{equation}

Exploiting the
relator in (\ref{H}) equations (\ref{Wna}) become
\begin{align*}
  &A_{\v{k}}=e^{-i\frac{k_y}{2}}B_{\v{k}}, \qquad
  A_{\v{k}}'=e^{i\frac{k_y}{2}}\sigma_z B_{\v{k}}\sigma_z ,\\
  &B_\v{k}\mathrel{\mathop:}=\left(e^{-i\frac{k_x}{2}}A_a+e^{-i\frac{k_y}{2}}A_b+e^{i\frac{k_y}{2}}A_{a^{-1}}+e^{i\frac{k_x}{2}}A_{b^{-1}}\right),
\end{align*}
where $k_x:=k_2+k_3$ and $k_y:=k_1$. Defining
\[V \mathrel{\mathop:}= \begin{pmatrix} I &0\\
  0&e^{i\frac{k_y}{2}}\sigma_z
\end{pmatrix},\qquad R:=\frac{1}{\sqrt{2}}\begin{pmatrix} 1 &1\\ 1&-1
\end{pmatrix},
\]
one has
\begin{equation}\label{renorm}
\mathcal{R}[A]_{\v{k}} = V(R\otimes I)\left(\sigma_z\otimes
  B_{\v{k}}\sigma_z \right)(R^{\dagger}\otimes I)V^{\dagger}.
\end{equation}
Accordingly $\mathcal{R}\left[A\right]_{\v{k}}$ is unitarily
equivalent to $\sigma_z\otimes B_{\v{k}}\sigma_z$ whose four
eigenvalues $e^{-i\omega_r^{\pm} (k_x,k_y)}$, $r=1,2$ are
expressed in terms of the walk dispersion relations
\begin{equation*}
\begin{aligned}
  & \omega^{\pm}_1 :=\pm\arccos\alpha(k_x,k_y)+\pi/2,\qquad \omega^{\pm}_2:=\omega^{\pm}_1+\pi,\\
  & \alpha(k_x,k_y) :=\frac{1}{2}\left(\sin\frac{k_x}{2}+\sin\frac{k_y}{2}\right).
\end{aligned}
\end{equation*}
We notice that this dispersion relations are equivalent, up to shifts
in wave-vectors, to the Weyl QW one. The Weyl QW (which is the unique
isotropic QW for $s=2$ on the Cayley graph
$\mathbb{Z}^2=\<a,b|aba^{-1}b^{-1}\>$ \cite{PhysRevA.90.062106}).
While the coarse-graining procedure for the example in Section
\ref{s:example1} led to a class of walks with ``massive'' dispersion
relation, in this case the dispersion relation close to its minimum
(for wave-vectors close to $(\pi,\pi)$) is linear in $\v{k}$. In this
sense the non-Abelianity of the original graph does not induce
relevant effects on the dynamics of the coarse-grained QW.

We conclude analyzing the relation between the coarse-graining of the
two solutions obtained in this section, namely
$\mathcal{R}[A^{\mathrm{I}}]_{\v{k}}$ and
$\mathcal{R}[A^{\mathrm{II}}]_{\v{k}}$. Through a block diagonal
change of basis matrix, we saw that $\mathcal{R}[A^{I}]_{\v{k}}$ is
unitarily equivalent to $\sigma_z\otimes B_{\v{k}}\sigma_z$. In Since
two solutions $A^{\I}$ and $A^{\II}$ are connected by a local
anti-unitary transformation (see Appendix \ref{app:b}),
$\mathcal{R}[A^{II}]_{\v{k}}$ is, by linearity, unitarily equivalent
to $\sigma_z\otimes \sigma_zB^T_{\v{k}}$, which is just
$\mathcal{R}[A^{I}]^{\dagger}_{-\v{k}}$ (up to a change of
basis). This means that the two coarse-grained QWs are connected by
$\mathsf{PT}$ symmetry, with parity and time-reversal maps given by
$\mathsf{P}: \v{k} \longmapsto -\v{k}$ and
$ \mathsf{T}: A\ \longmapsto A^{\dagger}$.

\section{Conclusions}

In this manuscript we considered quantum walks on Cayley graphs of
non-Abelian groups, with emphasis on the case of virtually Abelian
groups.  We devised a coarse-graining technique to reduce the quantum
walk to an equivalent one on an Abelian group with coin system of
larger dimension. This method, based on the group structure of the
graph, allows one to extend the notion of wave-vector as an invariant
of the walk dynamics to the virtually Abelian case. Within this
framework, virtually Abelian QWs can be diagonalized in terms of a
dispersion relation, which carries relevant kinematic information.

We derived the class of QWs allowed on two special virtually Abelian groups with planar Cayley graphs. We then applied the coarse-graining technique to solve the aforementioned walks
analytically. Interestingly, for one of the two groups the dynamics of the QW leads to a massive dispersion relation by virtue of the presence of nontrivial cyclic subgroups.

The coarse-graining technique represents a crucial step in
the derivation of the complete set of quantum walks allowed on Cayley graphs that are quasi-isometric to $\mathbb{R}^d$.

Finally, our technique can also be exploited as an intermediate step in a general renormalization procedure, where besides coarse-graining one also needs to implement the disposal of information for the purpose of changing the relevant scale of description of the walk dynamics. For example, this can be accomplished by an algebraic approach to renormalization as illustrated in Ref.~\cite{1367-2630-17-8-083005}.

\appendix

\acknowledgments
This work has been supported in part by the Templeton Foundation under the project ID\# 43796 {\em A
  Quantum-Digital Universe}. The authors would like to acknowledge valuable suggestions and discussions with Franco Manessi. 
  
\bibliography{bibliography2}

\begin{thebibliography}{22}%
\makeatletter
\providecommand \@ifxundefined [1]{%
 \@ifx{#1\undefined}
}%
\providecommand \@ifnum [1]{%
 \ifnum #1\expandafter \@firstoftwo
 \else \expandafter \@secondoftwo
 \fi
}%
\providecommand \@ifx [1]{%
 \ifx #1\expandafter \@firstoftwo
 \else \expandafter \@secondoftwo
 \fi
}%
\providecommand \natexlab [1]{#1}%
\providecommand \enquote  [1]{``#1''}%
\providecommand \bibnamefont  [1]{#1}%
\providecommand \bibfnamefont [1]{#1}%
\providecommand \citenamefont [1]{#1}%
\providecommand \href@noop [0]{\@secondoftwo}%
\providecommand \href [0]{\begingroup \@sanitize@url \@href}%
\providecommand \@href[1]{\@@startlink{#1}\@@href}%
\providecommand \@@href[1]{\endgroup#1\@@endlink}%
\providecommand \@sanitize@url [0]{\catcode `\\12\catcode `\$12\catcode
  `\&12\catcode `\#12\catcode `\^12\catcode `\_12\catcode `\%12\relax}%
\providecommand \@@startlink[1]{}%
\providecommand \@@endlink[0]{}%
\providecommand \url  [0]{\begingroup\@sanitize@url \@url }%
\providecommand \@url [1]{\endgroup\@href {#1}{\urlprefix }}%
\providecommand \urlprefix  [0]{URL }%
\providecommand \Eprint [0]{\href }%
\providecommand \doibase [0]{http://dx.doi.org/}%
\providecommand \selectlanguage [0]{\@gobble}%
\providecommand \bibinfo  [0]{\@secondoftwo}%
\providecommand \bibfield  [0]{\@secondoftwo}%
\providecommand \translation [1]{[#1]}%
\providecommand \BibitemOpen [0]{}%
\providecommand \bibitemStop [0]{}%
\providecommand \bibitemNoStop [0]{.\EOS\space}%
\providecommand \EOS [0]{\spacefactor3000\relax}%
\providecommand \BibitemShut  [1]{\csname bibitem#1\endcsname}%
\let\auto@bib@innerbib\@empty
\bibitem [{\citenamefont {Ambainis}\ \emph {et~al.}(2001)\citenamefont
  {Ambainis}, \citenamefont {Bach}, \citenamefont {Nayak}, \citenamefont
  {Vishwanath},\ and\ \citenamefont {Watrous}}]{ambainis2001one}%
  \BibitemOpen
  \bibfield  {author} {\bibinfo {author} {\bibfnamefont {A.}~\bibnamefont
  {Ambainis}}, \bibinfo {author} {\bibfnamefont {E.}~\bibnamefont {Bach}},
  \bibinfo {author} {\bibfnamefont {A.}~\bibnamefont {Nayak}}, \bibinfo
  {author} {\bibfnamefont {A.}~\bibnamefont {Vishwanath}}, \ and\ \bibinfo
  {author} {\bibfnamefont {J.}~\bibnamefont {Watrous}},\ }in\ \href {\doibase
  10.1145/380752.380757} {\emph {\bibinfo {booktitle} {Proceedings of the
  Thirty-third Annual ACM Symposium on Theory of Computing}}},\ \bibinfo
  {series and number} {STOC '01}\ (\bibinfo  {publisher} {ACM},\ \bibinfo
  {address} {New York, NY, USA},\ \bibinfo {year} {2001})\ pp.\ \bibinfo
  {pages} {37--49}\BibitemShut {NoStop}%
\bibitem [{\citenamefont {Aharonov}\ \emph {et~al.}(2001)\citenamefont
  {Aharonov}, \citenamefont {Ambainis}, \citenamefont {Kempe},\ and\
  \citenamefont {Vazirani}}]{aharonov2001quantum}%
  \BibitemOpen
  \bibfield  {author} {\bibinfo {author} {\bibfnamefont {D.}~\bibnamefont
  {Aharonov}}, \bibinfo {author} {\bibfnamefont {A.}~\bibnamefont {Ambainis}},
  \bibinfo {author} {\bibfnamefont {J.}~\bibnamefont {Kempe}}, \ and\ \bibinfo
  {author} {\bibfnamefont {U.}~\bibnamefont {Vazirani}},\ }in\ \href {\doibase
  10.1145/380752.380758} {\emph {\bibinfo {booktitle} {Proceedings of the
  Thirty-third Annual ACM Symposium on Theory of Computing}}},\ \bibinfo
  {series and number} {STOC '01}\ (\bibinfo  {publisher} {ACM},\ \bibinfo
  {address} {New York, NY, USA},\ \bibinfo {year} {2001})\ pp.\ \bibinfo
  {pages} {50--59}\BibitemShut {NoStop}%
\bibitem [{\citenamefont {Knight}\ \emph {et~al.}(2004)\citenamefont {Knight},
  \citenamefont {Rold\'an},\ and\ \citenamefont
  {Sipe}}]{doi:10.1080/09500340408232489}%
  \BibitemOpen
  \bibfield  {author} {\bibinfo {author} {\bibfnamefont {P.~L.}\ \bibnamefont
  {Knight}}, \bibinfo {author} {\bibfnamefont {E.}~\bibnamefont {Rold\'an}}, \
  and\ \bibinfo {author} {\bibfnamefont {J.~E.}\ \bibnamefont {Sipe}},\ }\href
  {\doibase 10.1080/09500340408232489} {\bibfield  {journal} {\bibinfo
  {journal} {Journal of Modern Optics}\ }\textbf {\bibinfo {volume} {51}},\
  \bibinfo {pages} {1761} (\bibinfo {year} {2004})}\BibitemShut {NoStop}%
\bibitem [{\citenamefont {D'Ariano}\ and\ \citenamefont
  {Perinotti}(2014)}]{PhysRevA.90.062106}%
  \BibitemOpen
  \bibfield  {author} {\bibinfo {author} {\bibfnamefont {G.~M.}\ \bibnamefont
  {D'Ariano}}\ and\ \bibinfo {author} {\bibfnamefont {P.}~\bibnamefont
  {Perinotti}},\ }\href {\doibase 10.1103/PhysRevA.90.062106} {\bibfield
  {journal} {\bibinfo  {journal} {Phys. Rev. A}\ }\textbf {\bibinfo {volume}
  {90}},\ \bibinfo {pages} {062106} (\bibinfo {year} {2014})}\BibitemShut
  {NoStop}%
\bibitem [{\citenamefont {Gromov}()}]{gromov}%
  \BibitemOpen
  \bibfield  {author} {\bibinfo {author} {\bibfnamefont {M.}~\bibnamefont
  {Gromov}},\ }\href {\doibase 10.1007/s000390300002} {\bibfield  {journal}
  {\bibinfo  {journal} {Geometric \& Functional Analysis GAFA}\ }\textbf
  {\bibinfo {volume} {13}},\ \bibinfo {pages} {73}}\BibitemShut {NoStop}%
\bibitem [{\citenamefont {Arrighi}\ \emph {et~al.}(2012)\citenamefont
  {Arrighi}, \citenamefont {Martiel},\ and\ \citenamefont
  {Nesme}}]{2012arXiv1212.0027A}%
  \BibitemOpen
  \bibfield  {author} {\bibinfo {author} {\bibfnamefont {P.}~\bibnamefont
  {Arrighi}}, \bibinfo {author} {\bibfnamefont {S.}~\bibnamefont {Martiel}}, \
  and\ \bibinfo {author} {\bibfnamefont {V.}~\bibnamefont {Nesme}},\ }\href
  {http://arxiv.org/abs/1212.0027} {\bibfield  {journal} {\bibinfo  {journal}
  {arXiv:1212.0027}\ } (\bibinfo {year} {2012})}\BibitemShut {NoStop}%
\bibitem [{\citenamefont {Acevedo}\ and\ \citenamefont
  {Gobron}(2006)}]{acevedo2006quantum}%
  \BibitemOpen
  \bibfield  {author} {\bibinfo {author} {\bibfnamefont {O.~L.}\ \bibnamefont
  {Acevedo}}\ and\ \bibinfo {author} {\bibfnamefont {T.}~\bibnamefont
  {Gobron}},\ }\href@noop {} {\bibfield  {journal} {\bibinfo  {journal}
  {Journal of Physics A: Mathematical and General}\ }\textbf {\bibinfo {volume}
  {39}},\ \bibinfo {pages} {585} (\bibinfo {year} {2006})}\BibitemShut
  {NoStop}%
\bibitem [{\citenamefont {Acevedo}\ \emph {et~al.}(2008)\citenamefont
  {Acevedo}, \citenamefont {Roland},\ and\ \citenamefont
  {Cerf}}]{Acevedo:2008:ESQ:2011752.2011757}%
  \BibitemOpen
  \bibfield  {author} {\bibinfo {author} {\bibfnamefont {O.~L.}\ \bibnamefont
  {Acevedo}}, \bibinfo {author} {\bibfnamefont {J.}~\bibnamefont {Roland}}, \
  and\ \bibinfo {author} {\bibfnamefont {N.~J.}\ \bibnamefont {Cerf}},\ }\href
  {http://dl.acm.org/citation.cfm?id=2011752.2011757} {\bibfield  {journal}
  {\bibinfo  {journal} {Quantum Info. Comput.}\ }\textbf {\bibinfo {volume}
  {8}},\ \bibinfo {pages} {68} (\bibinfo {year} {2008})}\BibitemShut {NoStop}%
\bibitem [{\citenamefont {Portugal}\ \emph
  {et~al.}(2015{\natexlab{a}})\citenamefont {Portugal}, \citenamefont {Santos},
  \citenamefont {Fernandes},\ and\ \citenamefont
  {Gon{\c{c}}alves}}]{Portugal2015}%
  \BibitemOpen
  \bibfield  {author} {\bibinfo {author} {\bibfnamefont {R.}~\bibnamefont
  {Portugal}}, \bibinfo {author} {\bibfnamefont {R.~A.~M.}\ \bibnamefont
  {Santos}}, \bibinfo {author} {\bibfnamefont {T.~D.}\ \bibnamefont
  {Fernandes}}, \ and\ \bibinfo {author} {\bibfnamefont {D.~N.}\ \bibnamefont
  {Gon{\c{c}}alves}},\ }\href@noop {} {\bibfield  {journal} {\bibinfo
  {journal} {Quantum Information Processing}\ }\textbf {\bibinfo {volume}
  {15}},\ \bibinfo {pages} {85} (\bibinfo {year}
  {2015}{\natexlab{a}})}\BibitemShut {NoStop}%
\bibitem [{\citenamefont {Portugal}\ \emph
  {et~al.}(2015{\natexlab{b}})\citenamefont {Portugal}, \citenamefont
  {Boettcher},\ and\ \citenamefont {Falkner}}]{PhysRevA.91.052319}%
  \BibitemOpen
  \bibfield  {author} {\bibinfo {author} {\bibfnamefont {R.}~\bibnamefont
  {Portugal}}, \bibinfo {author} {\bibfnamefont {S.}~\bibnamefont {Boettcher}},
  \ and\ \bibinfo {author} {\bibfnamefont {S.}~\bibnamefont {Falkner}},\ }\href
  {\doibase 10.1103/PhysRevA.91.052319} {\bibfield  {journal} {\bibinfo
  {journal} {Phys. Rev. A}\ }\textbf {\bibinfo {volume} {91}},\ \bibinfo
  {pages} {052319} (\bibinfo {year} {2015}{\natexlab{b}})}\BibitemShut
  {NoStop}%
\bibitem [{\citenamefont {Bisio}\ \emph {et~al.}(2015)\citenamefont {Bisio},
  \citenamefont {D'Ariano},\ and\ \citenamefont {Tosini}}]{Bisio2015244}%
  \BibitemOpen
  \bibfield  {author} {\bibinfo {author} {\bibfnamefont {A.}~\bibnamefont
  {Bisio}}, \bibinfo {author} {\bibfnamefont {G.~M.}\ \bibnamefont {D'Ariano}},
  \ and\ \bibinfo {author} {\bibfnamefont {A.}~\bibnamefont {Tosini}},\ }\href
  {\doibase http://dx.doi.org/10.1016/j.aop.2014.12.016} {\bibfield  {journal}
  {\bibinfo  {journal} {Annals of Physics}\ }\textbf {\bibinfo {volume}
  {354}},\ \bibinfo {pages} {244 } (\bibinfo {year} {2015})}\BibitemShut
  {NoStop}%
\bibitem [{\citenamefont {Bisio}\ \emph
  {et~al.}(2016{\natexlab{a}})\citenamefont {Bisio}, \citenamefont {D'Ariano},\
  and\ \citenamefont {Perinotti}}]{Bisio2016177}%
  \BibitemOpen
  \bibfield  {author} {\bibinfo {author} {\bibfnamefont {A.}~\bibnamefont
  {Bisio}}, \bibinfo {author} {\bibfnamefont {G.~M.}\ \bibnamefont {D'Ariano}},
  \ and\ \bibinfo {author} {\bibfnamefont {P.}~\bibnamefont {Perinotti}},\
  }\href {\doibase http://dx.doi.org/10.1016/j.aop.2016.02.009} {\bibfield
  {journal} {\bibinfo  {journal} {Annals of Physics}\ }\textbf {\bibinfo
  {volume} {368}},\ \bibinfo {pages} {177 } (\bibinfo {year}
  {2016}{\natexlab{a}})}\BibitemShut {NoStop}%
\bibitem [{\citenamefont {Bisio}\ \emph
  {et~al.}(2016{\natexlab{b}})\citenamefont {Bisio}, \citenamefont {D'Ariano},
  \citenamefont {Erba}, \citenamefont {Perinotti},\ and\ \citenamefont
  {Tosini}}]{PhysRevA.93.062334}%
  \BibitemOpen
  \bibfield  {author} {\bibinfo {author} {\bibfnamefont {A.}~\bibnamefont
  {Bisio}}, \bibinfo {author} {\bibfnamefont {G.~M.}\ \bibnamefont {D'Ariano}},
  \bibinfo {author} {\bibfnamefont {M.}~\bibnamefont {Erba}}, \bibinfo {author}
  {\bibfnamefont {P.}~\bibnamefont {Perinotti}}, \ and\ \bibinfo {author}
  {\bibfnamefont {A.}~\bibnamefont {Tosini}},\ }\href {\doibase
  10.1103/PhysRevA.93.062334} {\bibfield  {journal} {\bibinfo  {journal} {Phys.
  Rev. A}\ }\textbf {\bibinfo {volume} {93}},\ \bibinfo {pages} {062334}
  (\bibinfo {year} {2016}{\natexlab{b}})}\BibitemShut {NoStop}%
\bibitem [{\citenamefont {Kadanoff}(1966)}]{K66}%
  \BibitemOpen
  \bibfield  {author} {\bibinfo {author} {\bibfnamefont {L.~P.}\ \bibnamefont
  {Kadanoff}},\ }\href@noop {} {\bibfield  {journal} {\bibinfo  {journal}
  {Physics}\ }\textbf {\bibinfo {volume} {2}},\ \bibinfo {pages} {63} (\bibinfo
  {year} {1966})}\BibitemShut {NoStop}%
\bibitem [{\citenamefont {B{\'e}ny}\ and\ \citenamefont
  {Osborne}(2015)}]{1367-2630-17-8-083005}%
  \BibitemOpen
  \bibfield  {author} {\bibinfo {author} {\bibfnamefont {C.}~\bibnamefont
  {B{\'e}ny}}\ and\ \bibinfo {author} {\bibfnamefont {T.~J.}\ \bibnamefont
  {Osborne}},\ }\href {http://stacks.iop.org/1367-2630/17/i=8/a=083005}
  {\bibfield  {journal} {\bibinfo  {journal} {New Journal of Physics}\ }\textbf
  {\bibinfo {volume} {17}},\ \bibinfo {pages} {083005} (\bibinfo {year}
  {2015})}\BibitemShut {NoStop}%
\bibitem [{\citenamefont {Dehn}(1911)}]{dehn1911unendliche}%
  \BibitemOpen
  \bibfield  {author} {\bibinfo {author} {\bibfnamefont {M.}~\bibnamefont
  {Dehn}},\ }\href@noop {} {\bibfield  {journal} {\bibinfo  {journal}
  {Mathematische Annalen}\ }\textbf {\bibinfo {volume} {71}},\ \bibinfo {pages}
  {116} (\bibinfo {year} {1911})}\BibitemShut {NoStop}%
\bibitem [{\citenamefont {de~La~Harpe}(2000)}]{harpe}%
  \BibitemOpen
  \bibfield  {author} {\bibinfo {author} {\bibfnamefont {P.}~\bibnamefont
  {de~La~Harpe}},\ }\href@noop {} {\emph {\bibinfo {title} {Topics in geometric
  group theory}}}\ (\bibinfo  {publisher} {University of Chicago Press},\
  \bibinfo {year} {2000})\BibitemShut {NoStop}%
\bibitem [{\citenamefont {Gromov}(1984)}]{gromov1984infinite}%
  \BibitemOpen
  \bibfield  {author} {\bibinfo {author} {\bibfnamefont {M.}~\bibnamefont
  {Gromov}},\ }in\ \href@noop {} {\emph {\bibinfo {booktitle} {Proc.
  International Congress of Mathematicians}}},\ Vol.~\bibinfo {volume} {1}\
  (\bibinfo {year} {1984})\ p.~\bibinfo {pages} {2}\BibitemShut {NoStop}%
\bibitem [{\citenamefont {Campbell}(1999)}]{campbell1999groups}%
  \BibitemOpen
  \bibfield  {author} {\bibinfo {author} {\bibfnamefont {C.~M.}\ \bibnamefont
  {Campbell}},\ }\href {https://books.google.it/books?id=1UbpiaSYSNYC} {\emph
  {\bibinfo {title} {Groups St Andrews 1997 in Bath:}}},\ Groups St Andrews
  1997 in Bath; I-II: Proceedings\ (\bibinfo  {publisher} {Cambridge University
  Press},\ \bibinfo {year} {1999})\BibitemShut {NoStop}%
\bibitem [{\citenamefont {Kapovich}(2012)}]{Kapovich:2012te}%
  \BibitemOpen
  \bibfield  {author} {\bibinfo {author} {\bibfnamefont {M.}~\bibnamefont
  {Kapovich}},\ }\href@noop {} {\bibfield  {journal} {\bibinfo  {journal}
  {Publications of IAS/Park City Summer Institute}\ } (\bibinfo {year}
  {2012})}\BibitemShut {NoStop}%
\bibitem [{\citenamefont {de~Cornulier}\ \emph {et~al.}(2007)\citenamefont
  {de~Cornulier}, \citenamefont {Tessera},\ and\ \citenamefont
  {Valette}}]{rigidity}%
  \BibitemOpen
  \bibfield  {author} {\bibinfo {author} {\bibfnamefont {Y.}~\bibnamefont
  {de~Cornulier}}, \bibinfo {author} {\bibfnamefont {R.}~\bibnamefont
  {Tessera}}, \ and\ \bibinfo {author} {\bibfnamefont {A.}~\bibnamefont
  {Valette}},\ }\href {\doibase 10.1007/s00039-007-0604-0} {\bibfield
  {journal} {\bibinfo  {journal} {GAFA Geometric And Functional Analysis}\
  }\textbf {\bibinfo {volume} {17}},\ \bibinfo {pages} {770} (\bibinfo {year}
  {2007})}\BibitemShut {NoStop}%
\bibitem [{\citenamefont {D'Ariano}(2012)}]{mauro2012quantum}%
  \BibitemOpen
  \bibfield  {author} {\bibinfo {author} {\bibfnamefont {G.~M.}\ \bibnamefont
  {D'Ariano}},\ }\href {\doibase
  http://dx.doi.org/10.1016/j.physleta.2011.12.021} {\bibfield  {journal}
  {\bibinfo  {journal} {Physics Letters A}\ }\textbf {\bibinfo {volume}
  {376}},\ \bibinfo {pages} {697 } (\bibinfo {year} {2012})}\BibitemShut
  {NoStop}%
\end{thebibliography}%

\section{Derivation of the QW in Section \ref{s:example1}}\label{app:a}

Here we derive the walk in Section \ref{s:example1} on the Cayley
graph of $G_1 =\langle a,b\ |\ a^4,b^4,{\left(ab\right)}^{2}\rangle$.
Assuming the isotropy of the walk (see Section
\ref{s:qws-on-cayley}), we show how to solve the unitarity constraints
Eqs.~(\ref{eq:ort1},\ref{eq:sum1},\ref{eq:norm1}) obtained from the
above Cayley graph.
We take the polar decomposition\footnote{Every complex square matrix
  admits a so called \emph{polar decomposition}, namely
  $\forall M \in \mathcal{M}\left( 2 \times 2, \mathbb{C}\right)\
  \exists \ V \in \mathcal{M}\left( 2\times 2, \mathbb{C}\right)$
  unitary, $P \in \mathcal{M}\left( 2\times 2, \mathbb{C}\right)$
  semi-positive definite : $M=VP$.} of the QW transition matrices
$A_g=V_gP_g$ ($g\in S=\{a,b,a^{-1},b^{-1}\}$) and considering that
$P_q=P_q^{\dagger}$, the conditions in Eq.\eqref{eq:ort1} become
\begin{align}\label{eq:pol1}
V_{i^{\pm 1}}P_{i^{\pm 1}}{P_{{j}^{\pm 1}}}{V_{{j}^{\pm 1}}}^{\dagger} & = 0, \\\label{eq:pol2}
P_{i^{\pm 1}}{V_{i^{\pm 1}}}^{\dagger}{V_{{j}^{\pm 1}}}{P_{{j}^{\pm 1}}} & = 0.
\end{align}

Being the $V$ matrices unitary, we need to have $P_{i^{\pm 1}}{P_{{j}^{\pm 1}}}=0$, and being both $P_i$ and $P_j$ nonnull (all transition matrices are nonnull by definition) they must be both rank one.  Moreover, from \eqref{eq:pol1} one has
\begin{align*}
P_{i^{\pm 1}}=\alpha_{i^{\pm 1}}\ket{+_{i^{\pm 1}}}\bra{+_{i^{\pm 1}}},\ \ P_{j^{\pm 1}}=\alpha_{j^{\pm 1}}\ket{-_{i^{\pm 1}}}\bra{-_{i^{\pm 1}}},
\end{align*}
with $\alpha_g>0$ and $\left\{\ket{+_g},\ket{-_g} \right\}$
orthonormal bases for $\mathbb{C}^2$.  From equation \eqref{eq:pol2}
we get
$\bra{+_{i^{\pm 1}}}{V_{i^{\pm 1}}}^{\dagger}V_{{j}^{\pm
    1}}\ket{-_{i^{\pm 1}}}=0$
and, since ${V_{i^{\pm 1}}}^{\dagger}V_{{j}^{\pm 1}}$ is unitary, it
must be diagonal on the basis
$\left\lbrace\ket{+_{i^{\pm 1}}},\ket{-_{i^{\pm 1}}} \right\rbrace$
(with entries corresponding to phases).  The $V_g$ are not
uniquely determined by the polar decomposition, since the $A_g$ are
not full rank. In fact, if $A_{i^{\pm 1}}=V_{i^{\pm 1}}P_{i^{\pm 1}}$
holds for a $V_{i^{\pm 1}}$, there exists an infinite class of unitary
matrices $V_{i^{\pm 1}}'$ such that
$A_{i^{\pm 1}}=V_{i^{\pm 1}}'P_{i^{\pm 1}}$: all the unitary
matrices
\begin{align}
V_{i^{\pm 1}}'=V_{i^{\pm 1}}\left(\ket{+_{i^{\pm 1}}}\bra{+_{i^{\pm 1}}}+e^{i\theta_{i^{\pm 1}}}\ket{-_{i^{\pm 1}}}\bra{-_{i^{\pm 1}}}\right)
\end{align}
give the same polar decomposition for $A_{i^{\pm 1}}$. The same
freedom holds for $A_{{j}^{\pm 1}}$, $j\neq i$ and one can
always fix it computing
\[
\begin{split}
{V_{i^{\pm 1}}'}^{\dagger}V_{j^{\pm 1}}' & = e^{i\theta_{j^{\pm 1}}}\bra{+_{i^{\pm 1}}}{V_{i^{\pm 1}}}^{\dagger}V_{j^{\pm 1}}\ket{+_{i^{\pm 1}}}\ket{+_{i^{\pm 1}}}\bra{+_{i^{\pm 1}}} \\ &+e^{-i\theta_{i^{\pm 1}}}\bra{-_{i^{\pm 1}}}{V_{i^{\pm 1}}}^{\dagger}V_{{j}^{\pm 1}}\ket{-_{i^{\pm 1}}}\ket{-_{i^{\pm 1}}}\bra{-_{i^{\pm 1}}}
\end{split}
\]
and posing
$e^{i\theta_{j^{\pm 1}}}\bra{+_{i^{\pm 1}}}{V_{i^{\pm
      1}}}^{\dagger}V_{j^{\pm 1}}\ket{+_{i^{\pm 1}}}=1$
and
$e^{-i\theta_{i^{\pm 1}}}\bra{-_{i^{\pm 1}}}{V_{i^{\pm
      1}}}^{\dagger}V_{j^{\pm 1}}\ket{-_{i^{\pm 1}}}=1$.
Accordingly one has
\begin{equation}
V_{i^{\pm 1}}'^{\dagger}V_{{j}^{\pm 1}}'=I\Longrightarrow \ V_{{j}^{\pm 1}}'=V_{i^{\pm 1}}'
\end{equation}
that leads to the following structure for the transition matrices
\begin{align}\label{eq:pol-def}
A_a = \alpha_a V\ket{+_a}\bra{+_a}, A_b = \alpha_b V\ket{-_a}\bra{-_a}, \\
A_{a^{-1}} = \alpha_{a^{-1}} W\ket{+_{a^{-1}}}\bra{+_{a^{-1}}}, \\  A_{b^{-1}} = \alpha_{b^{-1}} W\ket{-_{a^{-1}}}\bra{-_{a^{-1}}},
\end{align}
with $\alpha_g>0$ and $\{\ket{+_q},\ket{-_q} \}$ orthonormal bases for $\mathbb{C}^2$.

Combining \eqref{eq:ort1} and \eqref{eq:sum1} one obtains
\begin{align*}
A_{i^{-1}}A_i^{\dagger}A_{j^{-1}}  = 0, \qquad
A_iA_{i^{-1}}^{\dagger}A_j  = 0, \\
A_jA_{i^{-1}}^{\dagger}A_i  = 0, \qquad
A_{j^{-1}}A_i^{\dagger}A_{i^{-1}}  = 0.
\end{align*}
and using the expression \eqref{eq:pol-def} for the $A_g$ we get
\begin{equation}\label{eq:sum-pol}
\begin{aligned}
\braket{+_a}{+_{a^{-1}}} \bra{-_{a^{-1}}}W^{\dagger}V\ket{+_a}  = 0, \\
\braket{-_a}{-_{a^{-1}}} \bra{+_{a^{-1}}}W^{\dagger}V\ket{-_a}  = 0, \\
\braket{+_a}{-_{a^{-1}}} \bra{+_{a^{-1}}}W^{\dagger}V\ket{+_a}  = 0, \\
\braket{-_a}{+_{a^{-1}}} \bra{-_{a^{-1}}}W^{\dagger}V\ket{-_a}  = 0. 
\end{aligned}
\end{equation}
Considering that $\braket{+_a}{-_{a^{-1}}}=0\Leftrightarrow
\ket{+_a}=\ket{+_{a^{-1}}} \Leftrightarrow \braket{-_a}{+_{a^{-1}}}=0$
(up to phase factors that would not appear in the $A_g$), condition
\eqref{eq:sum-pol} can be satisfied only in two cases

\begin{enumerate}[I]
\item
\hfill$\begin{aligned}[t]
\bra{+_{a^{-1}}}W^{\dagger}V\ket{-_a}=\bra{-_{a^{-1}}}W^{\dagger}V\ket{+_a}=0, \\
\end{aligned}$\hfill\null
\item
\hfill$\begin{aligned}[t]
 \bra{-_{a^{-1}}}W^{\dagger}V\ket{-_a} =\bra{+_{a^{-1}}}W^{\dagger}V\ket{+_a}=0.
\end{aligned}$\hfill\null
\end{enumerate}

Let's note that just two of the matrix elements which appear in (\ref{ort1}) can be zero: indeed, suppose by contradiction that this is not be the case, let's define $U$ any of the possible matrices which connect the two orthonormal bases found; thus $U^{\dagger}W^{\dagger}V$ would have at least three vanishing matrix elements, but this is absurd for it is unitary.
Accordingly, the two cases are:

\begin{enumerate}[I]
\item \label{eq:case1-1}
\hfill$\begin{aligned}[t]
A^{\I}_a  &= \alpha_a \mu W \ket{+}\bra{+} &
A^{\I}_b &= \alpha_b \nu W \ket{-}\bra{-} \\
A^{\I}_{a^{-1}}  &= \alpha_{a^{-1}} W \ket{+}\bra{+} &
A^{\I}_{b^{-1}}  &= \alpha_{b^{-1}} W \ket{-}\bra{-},
\end{aligned}$\hfill\null
\item \label{eq:case2-1}
\hfill$\begin{aligned}[t]
A^{\II}_a & = \alpha_a \mu W \ket{+}\bra{+} &
A^{\II}_b  &= \alpha_b \nu W \ket{-}\bra{-} \\
A^{\II}_{a^{-1}} & = \alpha_{a^{-1}} W \ket{-}\bra{-} &
A^{\II}_{b^{-1}}  &= \alpha_{b^{-1}} W \ket{+}\bra{+},
\end{aligned}$\hfill\null
\end{enumerate}
with $\mu,\nu$ phase factors and each of them can be equal either to
$i$ or to $-i$.  From the condition (\ref{eq:isotropy2}), $W$
is found simply substituting the $A_g$ and inverting the resulting
relation, while from the normalization (\ref{eq:norm1}) one can
find the $\alpha_g$.  The transition matrices for the case \ref{eq:case1-1} are
\begin{align*}
A^{\I}_a  &= \zeta^{\pm} \begin{pmatrix}
  1       &   0  \\
  0    &   0\end{pmatrix},         &  A_b &= \zeta^{\pm} \begin{pmatrix}
  0       &   0  \\
  0    &   1\end{pmatrix},       \\
A^{\I}_{a^{-1}} &= \zeta^{\mp}  \begin{pmatrix}
  1    &   0  \\
  0    &   0  \end{pmatrix},      &  A^{\I}_{b^{-1}}&= \zeta^{\mp}  \begin{pmatrix}
  0    &  0 \\
  0    &  1  \end{pmatrix}.
\end{align*}
where $A^{\I}_{a^{-1}}= (A^{\I}_a)^{\dagger}$, $A^{\I}_{b^{-1}}=(A^{\I}_b)^{\dagger}$, $\zeta^{\pm}\mathrel{\mathop:}=\frac{1 \pm i}{2}$, while the
solutions \ref{eq:case2-1} are the same up to the swap
$a^{-1}\leftrightarrow b^{-1}$.  

It is now simple to verify that the solutions derived in this section
satisfy the isotropy constraint \eqref{eq:isotropy}. Indeed for the
group $G_1:=\langle a,b\ |\ a^4,b^4,{\left(ab\right)}^{2}\rangle$ the
only transitive automorphism of $S_+=\{a,b\}$ is the swap
$a \leftrightarrow b$ of the generators. It is easy to verify that this automorphism is represented by
the unitary matrix $\sigma_x$.

Accordingly we see that by left multiplication of the transition
matrices with a unitary matrix commuting with $\sigma_x$---whose
general form is
\begin{align*}
Z_{\pm}=n I\pm im\sigma_x,
\end{align*}
for $n,m\geq 0$ and $n^2+m^2=1$---both the unitarity conditions and
the isotropy of the QW are unchanged. This shows that whole class of
isotropic QWs on $G_1$ is obtained by left multiplication of 
the above solutions by the matrix $Z_{\pm}=n I\pm im\sigma_x$.

\section{Derivation of the QW in Section \ref{s:example2}}\label{app:b}

In this appendix we present the derivation of the walk in Section \ref{s:example2}
on the Cayley graph of $G_2 = \langle a,b\ |\ a^2b^{-2}\rangle$.
We know that this corresponds to solve the unitarity conditions
in Eqs.~(\ref{eq:ort2},\ref{eq:sum2},\ref{eq:norm2}) for
the walk transition matrices. The procedure is very similar to the
case analysed in the previous section but for the convenience of the
reader we detail the full derivation.

Taking a polar decomposition $A_g=V_gP_g$ for the transition
matrices and noticing that $P_g=P_g^{\dagger}$, Eq.~\eqref{eq:ort2} becomes
\begin{align}
V_iP_i{P_{{j}^{-1}}}{V_{{j}^{-1}}}^{\dagger} & = 0, \label{6}\\
P_i{V_i}^{\dagger}{V_{{j}^{-1}}}{P_{{j}^{-1}}} & = 0, \label{7}
\end{align}
for $i\neq j\in\{a,b\}$. Being the $V$ matrices unitary, we need to have $P_{i^{\pm 1}}{P_{{j}^{\pm 1}}}=0$, and being both $P_i$ and $P_{j^{-1}}$ nonnull (all transition matrices are nonnull by definition) they must be both rank one.  Furthermore from (\ref{6}) one can write
\[
P_i=\alpha_i\ket{+_i}\bra{+_i},\ \ P_{j^{-1}}=\alpha_{j^{-1}}\ket{-_i}\bra{-_i}
\]
with $\alpha_q>0$ and $\left\lbrace\ket{+_i},\ket{-_i} \right\rbrace$ orthonormal bases for $\mathbb{C}^2$.\\

From equations (\ref{7}) one finds that
$\bra{+_i}{V_i}^{\dagger}V_{{j}^{-1}}\ket{-_i}=0$ and, since
${V_i}^{\dagger}V_{{j}^{-1}}$ is unitary, it must be diagonal on the
basis $\left\lbrace\ket{+_i},\ket{-_i} \right\rbrace$ (with entries
given by phase factors). The $V_q$ are not uniquely determined by the
polar decomposition, since the $A_q$ are not full rank. In fact, if
$A_i=V_iP_i$ holds for a $V_i$, there exists an infinite class of
unitary matrices $V_i'$ such that $A_i=V_i'P_i$:
\[
V_i'=V_i\left(\ket{+_i}\bra{+_i}+e^{i\theta_i}\ket{-_i}\bra{-_i}\right)
\]
are unitary and give the $A_i$ polar decomposition for. The same holds
for $A_{{j}^{-1}}$ and one can always fix this arbitrariness computing
\begin{align}
{V_i'}^{\dagger}V_{j^{-1}}'=&e^{i\theta_{j^{-1}}}\bra{+_i}{V_i}^{\dagger}V_{j^{-1}}\ket{+_i}\ket{+_i}\bra{+_i}+\\
&e^{-i\theta_i}\bra{-_i}{V_i}^{\dagger}V_{{j}^{-1}}\ket{-_i}\ket{-_i}\bra{-_i}
\end{align}
and taking
$e^{i\theta_{j^{-1}}}\bra{+_i}{V_i}^{\dagger}V_{j^{-1}}\ket{+_i}=1$
and $e^{-i\theta_i}\bra{-_i}{V_i}^{\dagger}V_{j^{-1}}\ket{-_i}=1$. It
follows ${V_i'}^{\dagger}V_{{j}^{-1}}'=I_2\ \Longrightarrow \ V_{{j}^{-1}}'=V_i'$
that allows to write the transition matrices as
\begin{equation}
\begin{aligned}\label{eq:matr}
A_a & = \alpha_a V\ket{+_a}\bra{+_a} ,\ \ A_{b^{-1}} = \alpha_{b^{-1}} V\ket{-_a}\bra{-_a}, \\
A_b & = \alpha_b W\ket{+_b}\bra{+_b} ,\ \ A_{a^{-1}} = \alpha_{a^{-1}} W\ket{-_b}\bra{-_b}.
\end{aligned}
\end{equation}
Combining Eq.~\eqref{eq:sum2} and Eq.~\eqref{eq:ort2} we get
\begin{align*}
A_iA_j^{\dagger}A_i & = 0, &
A_iA_{i^{-1}}^{\dagger}A_i & = 0, \\
A_{i^{-1}}A_{j^{-1}}^{\dagger}A_{i^{-1}} & = 0, &
A_{i^{-1}}A_i^{\dagger}A_{i^{-1}} & = 0.
\end{align*}
that using the expressions in \eqref{eq:matr} are equivalent to
\begin{equation}\label{ort1}
\begin{aligned}
\braket{+_a}{-_b} \bra{-_b}W^{\dagger}V\ket{+_a} & = 0, \\
\braket{-_a}{+_b} \bra{+_b}W^{\dagger}V\ket{-_a} & = 0, \\
\braket{+_a}{+_b} \bra{+_b}W^{\dagger}V\ket{+_a} & = 0, \\
\braket{-_a}{-_b} \bra{-_b}W^{\dagger}V\ket{-_a} & = 0. 
\end{aligned}
\end{equation}
Considering that
$\braket{+_a}{-_b}=0\Leftrightarrow \ket{+_a}=\ket{+_b}
\Leftrightarrow \braket{-_a}{+_b}=0$
(up to phase factors that would not appear in the $A_g$), 
(\ref{ort1}) are satisfied in the following two cases\begin{enumerate}[I]
\item \hfill$\begin{aligned}[t]
\ket{+_a}  & = \ket{+_b} \mathrel{\mathop:}= \ket{+} \\
\ket{-_a}  & = \ket{-_b} \mathrel{\mathop:}= \ket{-} 
\end{aligned}$\hfill\null
\item 
\hfill$\begin{aligned}[t]
\ket{+_a}  & = \ket{-_b} \mathrel{\mathop:}= \ket{+} \\
\ket{-_a}  & = \ket{+_b} \mathrel{\mathop:}= \ket{-}
\end{aligned}$\hfill\null
\end{enumerate}

We observe that just two of the matrix elements which appear in
(\ref{ort1}) can be zero: indeed, suppose by contradiction that this is not
the case, let's denote by $U$ any of the possible matrices
connecting the two orthonormal bases; thus $U^{\dagger}W^{\dagger}V$
would have at least three vanishing entries, but this is absurd
because it is a $2\times 2$ unitary matrix. Accordingly, in both cases it is $V\mathrel{\mathop:}=W\left( \mu
  \ket{+}\bra{-} + \nu \ket{-}\bra{+}  \right)$ where, using Eq.~\eqref{eq:sum2}, one finds
\begin{equation}
 \begin{aligned}
\mu \nu =-1,\ \alpha_{a^{-1}}=\alpha_{b}\mathrel{\mathop:}=\beta,\ \alpha_{a}=\alpha_{b^{-1}}\mathrel{\mathop:}=\alpha,
\end{aligned}
\end{equation}
and 
\begin{enumerate}[I]
\item  \label{ca}
\hfill$\begin{aligned}[t]
A_a & = \alpha \nu W \ket{-}\bra{+} &
A_b & = \beta W \ket{+}\bra{+} \\
A_{a^{-1}} & = \beta W \ket{-}\bra{-} &
A_{b^{-1}} & = -\alpha \nu^{*} W \ket{+}\bra{-},
\end{aligned}$\hfill\null
\item  \label{cb}
\hfill$\begin{aligned}[t]
A_a & = \alpha \nu W \ket{-}\bra{+} &
A_b & = \beta W \ket{-}\bra{-} \\
A_{a^{-1}} & = \beta W \ket{+}\bra{+} &
A_{b^{-1}} & = -\alpha \nu^{*} W \ket{+}\bra{-},
\end{aligned}$\hfill\null
\end{enumerate}
with the two solutions simply connected through the swap $b\leftrightarrow a^{-1}$.

From the normalization condition (\ref{eq:norm2}) it straightforwardly
follows that $\beta = \sqrt{1-\alpha^2}$, while from the
condition (\ref{eq:isotropy2}), $W$ is found substituting the
$A_g$ and inverting the resulting relation. The transition matrices for case \ref{ca} are then
\begin{align*}
  A_a^{\mathrm{I}}  &= \alpha \begin{pmatrix}
  \alpha       &   0  \\
  \sqrt{1-\alpha^2} \nu    &   0\end{pmatrix},         \\ A_b^{\mathrm{I}} &= \sqrt{1-\alpha^2} \begin{pmatrix}
  \alpha       &   0  \\
  -\alpha \nu    &   0\end{pmatrix},       \\
A_{a^{-1}}^{\mathrm{I}} &= \sqrt{1-\alpha^2}  \begin{pmatrix}
  0       &   \alpha \nu^*  \\
  0    &   \alpha  \end{pmatrix},   \\  A_{b^{-1}}^{\mathrm{I}} &= \alpha  \begin{pmatrix}
  0       &  - \sqrt{1-\alpha^2} \nu^*  \\
  0    &   \alpha  \end{pmatrix}.
\end{align*}
As one can easily verify, a unitary matrix $X$ such that
$XA_a^{\ref{ca}}X^{\dagger}=A_b^{\ref{ca}}$ 
must be diagonal and so it would multiply the entries of
$A_a^{\ref{ca}}$ 
by a phase factor; accordingly, the transitive action
\eqref{eq:isotropy} of the isotropy 
group on $S_+=\{a,b\}$ imposes $\alpha^2 = 1-\alpha^2$, namely $\alpha = \frac{1}{\sqrt{2}}$.

Let's define the unitary $U\left( \nu\right) \mathrel{\mathop:}= \bigl( \begin{smallmatrix}
1&0\\ 0&\nu
\end{smallmatrix} \bigr)$:
and by direct computation verify that
\[
U\left( \nu\right) A_q\left( \nu \right) {U\left(
    \nu\right)}^{\dagger} = A_q\left( 1 \right).
\]
By linearity of the walk, this implies that with a local unitary
conjugation one can remove the dependence from the phase factor $\nu$
and the transition matrices \ref{ca} are finally given by
\begin{align*}
A_a^{\mathrm{I}}  &= \frac{1}{2} \begin{pmatrix}
  1       &   0  \\
  1    &   0\end{pmatrix},         &  A_b^{\mathrm{I}} &= \frac{1}{2} \begin{pmatrix}
  1       &   0  \\
  -1    &   0\end{pmatrix},       \\
A_{a^{-1}}^{\mathrm{I}} &= \frac{1}{2}  \begin{pmatrix}
  0       &   1  \\
  0    &   1  \end{pmatrix},      &  A_{b^{-1}}^{\mathrm{I}} &= \frac{1}{2}  \begin{pmatrix}
  0       &  - 1 \\
  0    &   1  \end{pmatrix}.
\end{align*}

Now we notice that the solution \ref{ca} and \ref{cb} are connected
by an anti-unitary transformation
\begin{equation*}
  Y{\left( A_g^{\mathrm{I}}\right)}^T Y^{\dagger}=A_g^{\mathrm{II}},\ \ \
  Y=\frac{1}{\sqrt{2}}\left( I_2 + i\sigma_y \right).
\end{equation*}

As in the previous example it is now simple to verify the isotropy of
the QWs derived in this section. The only $G_2$-automorphism
transitive on $S_+=\{a,b\}$ is the swap
$a \leftrightarrow b$ of the generators that can be
represented on the coin system $\mathbb{C}^2$ by
the unitary matrix $\sigma_z$ to satisfy Eq.~\eqref{eq:isotropy}. We could then multiply the transition matrices by an arbitrary unitary matrix commuting with $\sigma_z$, but from Eq. \eqref{renorm} it's easy to see that this just amounts to a shift in the $k_x,k_y$.

\end{document}